\documentclass[12pt]{article}

\usepackage{vmargin}
\setmarginsrb{3cm}{2cm}{3cm}{2cm}{1cm}{1cm}{1cm}{1cm}
\usepackage{amsmath}
\usepackage{amssymb}
\usepackage[dvips]{graphicx}
\usepackage{rotating}
\usepackage{psfrag}
\usepackage{fancyhdr}
\newcommand{\n}{\nonumber}

\newcommand{\be}{\nopagebreak[3]\begin{equation}}
\newcommand{\ee}{\end{equation}}
\newcommand{\ba}{\nopagebreak[3]\begin{eqnarray}}
\newcommand{\ea}{\end{eqnarray}}

\newcommand{\va}{\scriptscriptstyle}

\begin{document}
\title{The $SU(2)$ Black Hole entropy revisited}

\author{J. Engle\\
Institut f\"{u}r Theoretische Physik III,\\
Universit\"{a}t Erlangen-N\"{u}rnberg,\\
Staudtstra\ss e 7, 91058 Erlangen, Germany. \\ \\
K. Noui\\
Laboratoire de Math\'ematiques et Physique 
Th\'eorique\footnote{F\'ed\'eration Denis Poisson Orl\'eans-Tours,
CNRS/UMR 6083},\\
Parc de Grammont, 37200 Tours, France. \\ \\
A. Perez and D. Pranzetti\\
Centre de Physique Th\'eorique\footnote{Unit\'e
Mixte de Recherche (UMR 6207) du CNRS et des Universit\'es
Aix-Marseille I, Aix-Marseille II, et du Sud Toulon-Var; laboratoire
afili\'e \`a la FRUMAM (FR 2291)}, \\
Campus de Luminy, 13288
Marseille, France.}

\sloppy
\maketitle

\pagestyle{plain}
\date{}

\begin{abstract}
We study the state-counting problem that arises in the $SU(2)$ black hole entropy calculation in loop quantum gravity.
More precisely, we compute the leading term and the logarithmic correction of both the spherically symmetric and
the distorted $SU(2)$ black holes. Contrary to what has been done in previous works, we have to take into account
``quantum corrections'' in our framework  in the sense that the level $k$ of the Chern-Simons theory which describes
the black hole is finite and not sent to infinity. Therefore,
the new results presented here allow for the computation of the entropy in models where the quantum group corrections are 
important.
\end{abstract}

\sloppy
\maketitle

\subsection*{Introduction}

The black hole entropy calculation in the framework of loop quantum gravity \cite{lqg} is based on the effective description of the
quantum gravitational degrees of freedom at the black hole horizon obtained from a suitable quantization of the classical phase space describing isolated horizons (see \cite{ih} and references therein). In these models the degrees of freedom at the horizon are described by Chern-Simons theories  with $SU(2)$ (or $U(1)$) structure groups \cite{ab,jon, su21,su22,su23}.
The simplest models are those where spherical symmetry is imposed already at the classical level. In this case it is natural (although not necessary) to consider
$SU(2)$ (or $U(1)$) Chern-Simons theory with a level that scales with the macroscopic classical area $k\propto a_{\va H}$. This  makes
the state-counting (necessary for the computation of the entropy) a combinatorial problem  which can be entirely formulated in terms of the representation theory of the classical group $SU(2)$ (or $U(1)$): for practical purposes one can take $k=\infty$ from the starting point \cite{counting1, counting2, counting3, hanno}.

However, the perspective considered above can be completely changed if one studies the models in the recently introduced $SU(2)$ Chern-Simons formulation \cite{su21,su22,su23}. The necessity of an $SU(2)$ gauge invariant formulation comes from the requirement that the isolated horizon quantum constraints be consistently imposed in the quantum theory (in \cite{su21} it is shown how the $U(1)$ treatment leads to an artificially larger entropy due to the fact that some of the second class constraints arising from the $SU(2)$-to-$U(1)$ gauge fixing can only be imposed weakly). However, the $SU(2)$ formulation is not unique as  there is a one parameter family of classically equivalent $SU(2)$ connections parametrizations of the horizon degrees of freedom. More precisely,  in the passage from Palatini-like variables to connection variables that is necessary for the description of the horizon degrees of freedom in terms of Chern-Simons theory (central for the quantization), an ambiguity parameter arises \cite{su22,su23}. This is completely analogous to the situation in the bulk  where the Immirzi parameter reflects an ambiguity in the choice of $SU(2)$ variables  in the passage from Palatini variables to Ashtekar-Barbero connections (central for the quantization in the loop quantum gravity approach). In the case of the parametrization of the isolated horizon degrees of freedom, this ambiguity can be encoded in the value of the Chern-Simons level $k$, which, in addition to the Immirzi parameter, becomes an independent free parameter of the classical formulation of the isolated horizon-bulk system.

Therefore, it is no longer natural (nor necessary) to take $k\propto a_{\va H}$. On the contrary, it seems more natural to exploit the existence of this ambiguity by letting $k$ be arbitrary \footnote{It is a good thing that the effective treatment contains a free parameter arising at the boundary from exactly the analogous reason as the Immirzi parameter in the bulk parametrization of the phase space. This keeps open the possibility  that dynamical considerations  could lead to cancelation of both ambiguities producing Immirzi parameter independent predictions.}. In this way the $SU(2)$ classical representation theory involved in previous calculations should be replaced by the  representation theory of the quantum group $U_q(su(2))$ 
with $q$ a non-trivial root of unity. Thus quantum group corrections become central for the state-counting problem. In this paper
we study the finite $k$ counting problem by means of simple asymptotic methods.
The powerful methods that have been developed for the resolution of the counting problem in the $k=\infty$ \cite{counting3,hanno} are perhaps generalizable to the finite $k$ case. Here we follow a less rigorous and more physical approach. The formulation is partly inspired from a combination of ideas stemming from different
calculations in the literature \cite{counting2, kaulma, livi}.

\medskip

In the first section, we review some basic facts concerning the quantization of $SU(2)$ Chern-Simons theory
whose physical states are built from the representation theory of the quantum group $U_q(su(2))$ when $q$ is a root
of unity. For that reason, we recall some properties of the representation theory of $U_q(su(2))$,
which allows us to compute the dimension of the Chern-Simons theory Hilbert space ${\cal H}_{CS}$ when the space is a punctured 
two-sphere. In the second Section, we give a new integral formulation of the dimension of ${\cal H}_{CS}$ which appear to be much more
convenient to compute black hole entropy. In the last Section, we compute the leading term of the $SU(2)$-black hole entropy and its logarithmic 
corrections first for the spherically symmetric black hole and then for the distorted black hole. 
We adapt the techniques used in \cite{counting3} and firstly introduced in \cite{counting1} to compute the entropy
of a black hole. We are not going into the mathematical details of these techniques which has been very well exposed
in \cite{counting3} and which  are in fact very well-known in the domain of probabilities and used
to understand some properties of random walks. We recover that in the spherically symmetric and distorted black holes, the leading
term of the entropy is proportional to the area and the first corrections are still logarithmic: $S(a) \sim \alpha a + \beta \log a$. 
In the spherically symmetric case, $\alpha$ depends on the level $k$ and reaches the value obtained in previous calculations when $k$
goes to infinity; concerning $\beta$ it is independent of $k$ and is given by the value $-3/2$ as expected.
In the distorted case, $\alpha$ grows logarithmically with $k$ and $\beta$ is fixed to the value $-3$. 
We finish with a discussion.

\subsection*{1. The Chern-Simons Hilbert Space}
The Chern-Simons theory associated to the group $SU(2)$ is a gauge theory on a three dimensional manifold $M$
governed by the action
\begin{eqnarray}
S_k[A] \; = \; \frac{k}{4\pi} \int_{M} \langle A \wedge dA + \frac{2}{3} A \wedge A \wedge A \rangle
\end{eqnarray}
where $k$ is called the level of the action, $A$ is the local $SU(2)$-connection field and $\langle \cdot,\cdot \rangle$
is a notation for the $\mathfrak{su}(2)$ Killing form.
The Chern-Simons theory became really important when first it was shown to be closely related to three dimensional gravity \cite{AT,Witten1}
and above all when Witten showed \cite{Witten2} its amazing relation to manifold and knots invariants. Indeed, the Chern-Simons path integral
is a manifold invariant whereas the mean values of quantum observables naturally lead to Jones polynomials. For all these reasons, Chern-Simons
theory has been the center of a lot of interests and its quantization is now very well-known when the gauge group is compact, and in
particular when the gauge group is $SU(2)$.

\medskip

The covariant (path integral) and canonical quantizations offer the two main strategies to quantize the Chern-Simons
theory. These approaches are complementary: the covariant quantization leads easily to the fact that the level $k$ must be
an integer when the gauge group is compact \cite{Witten2}; the canonical quantization leads to a precise description of the Hilbert space
when the gauge group is compact but not only (see \cite{CK} for an introduction of the combinatorial quantization for example).
Both quantizations are necessary to understand how the mean value of
Wilson loops observables are related to knots polynomials like the (colored) Jones polynomial or its generalizations.

\medskip

Here we are exclusively interested in the description of the Hilbert space of Chern-Simons theory when the space
is a two-sphere punctured with a number $p$ of particles. At the classical level, each puncture, labelled by $\ell \in [1,p]$,
comes with an unitary irreducible representation $j_\ell$ of the gauge group $SU(2)$. At the quantum level, one shows that
the classical group gauge symmetry is replaced by a quantum group symmetry and the Hilbert space is constructed from
the representation theory of the quantum group $U_q(su(2))$ where $q=\exp(i\pi/(k+2))$ is necessarily a root of unity.
An immediate consequence is that the $SU(2)$ representations labeling the classical punctures become $U_q(su(2))$
representations which concretely implies  a cut-off on the punctures' representations which cannot be higher than $k/2$.
Then, the associated Hilbert space is the vector space
\be\label{Hilbert} H_k(j_1,\cdots,j_p) \; =  \; \text{Inv}(\otimes_\ell j_\ell)\ee
of invariant tensors in the tensor product $\otimes_\ell j_\ell$
of $U_q(su(2))$ representations endowed with a Hilbert structure defined, as in the classical case, from the Haar measure on the quantum group.
However, we will not be interested in the Hilbert structure of $H_k(j_1,\cdots,j_\ell)$ in the rest of the paper but rather in its
vector space structure and more precisely in its dimension. Indeed, the computation of the $SU(2)$ black hole entropy in Loop
Quantum Gravity needs  to be done precisely the calculation of the dimension of the previous vector space.

The calculation of the Hilbert space dimension makes use of the Verlinde coefficients.
In order to introduce these coefficients, we start by recalling some basic facts concerning the representation theory
of $U_q(su(2))$.

\subsubsection*{1.1. Basics of the representation theory of $U_q(su(2))$}
This Section is devoted to recall some basic results on the quantum group $U_q(su(2))$ we will need in the sequel.
We are not going to give a precise definition of this quantum group and a complete description of its properties.
Furthermore we will be interested only on some aspects concerning its representations theory and its recoupling theory.

The (standard) unitary irreducible representations of $U_q(su(2))$ are labelled by integers $j \leq k/2$.
The dimension $d_j$ of the $j$-representation is the same as in the
classical theory and then we have $d_j=2j+1$. Given an element $\xi \in U_q(su(2))$, its representation $\pi_j(\xi)$ is an endomorphism of the vector space
$V_j$.
Many formulae coming from the representation theory of $U_q(su(2))$ are expressed in terms of q-numbers $[x]$ defined for any complex
number $x$ by the relation:
$$
[x] \; = \; \frac{q^x-q^{-x}}{q-q^{-1}} \; = \; \frac{\sin(\frac{\pi}{k+2}x)}{\sin(\frac{\pi}{k+2})}\;.
$$

\begin{figure}[ht]
\psfrag{j1}{$j_1$}
\psfrag{j2}{$j_2$}
\psfrag{j3}{$j_3$}
\centering
\includegraphics[scale=0.3]{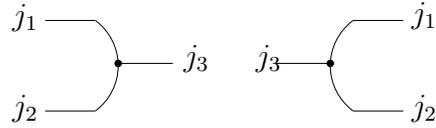}
\caption{Pictorial representation of the 3-valent intertwiner $\iota(j_1,j_2;j_3)$ and its adjoint operator $\iota(j_3;j_1,j_2)$.}
\label{3valent}
\end{figure}

Invariant $U_q(su(2))$-tensors are defined, by analogy with the classical situation, as tensors which are
invariant under the adjoint action. Note however that the adjoint
action is deformed compared to the classical case and makes use of the antipode instead of the inverse.
Among the invariant tensors, the 3-valent ones
$$\iota(j_1,j_2;j_3):  V_{j_1} \otimes V_{j_2} \; \longrightarrow \; V_{j_3}$$
are particularly interesting because all invariant tensors decompose into 3-valent intertwiners.
Three-valent intertwiners are represented as usual by a vertex between three lines colored by the representations
$j_\ell$ as illustrated in the figure (\ref{3valent}).
Contrary to what happens in the classical case where one can make a certain choice of normalization
such that the matrix elements of $\iota(j_1,j_2;j_3)$ are invariant under the permutations of $(j_1,j_2)$,
the order between the representations in $\iota(j_1,j_2,j_3)$ does
matter because of the presence of a non-trivial braiding in the quantum case.

\begin{figure}[ht]
\psfrag{j1}{$j_1$}
\psfrag{j2}{$j_2$}
\psfrag{j3}{$j_3$}
\psfrag{=}{$=$}
\psfrag{S}{$\sum_{j_3}[d_{j_3}]$}
\psfrag{C}{$Y(j_1,j_2,j_3)$}
\psfrag{A}{${Y(j_1,j_2,j_3)}{[d_{j_3}]}^{-1}$}
\centering
\includegraphics[scale=0.5]{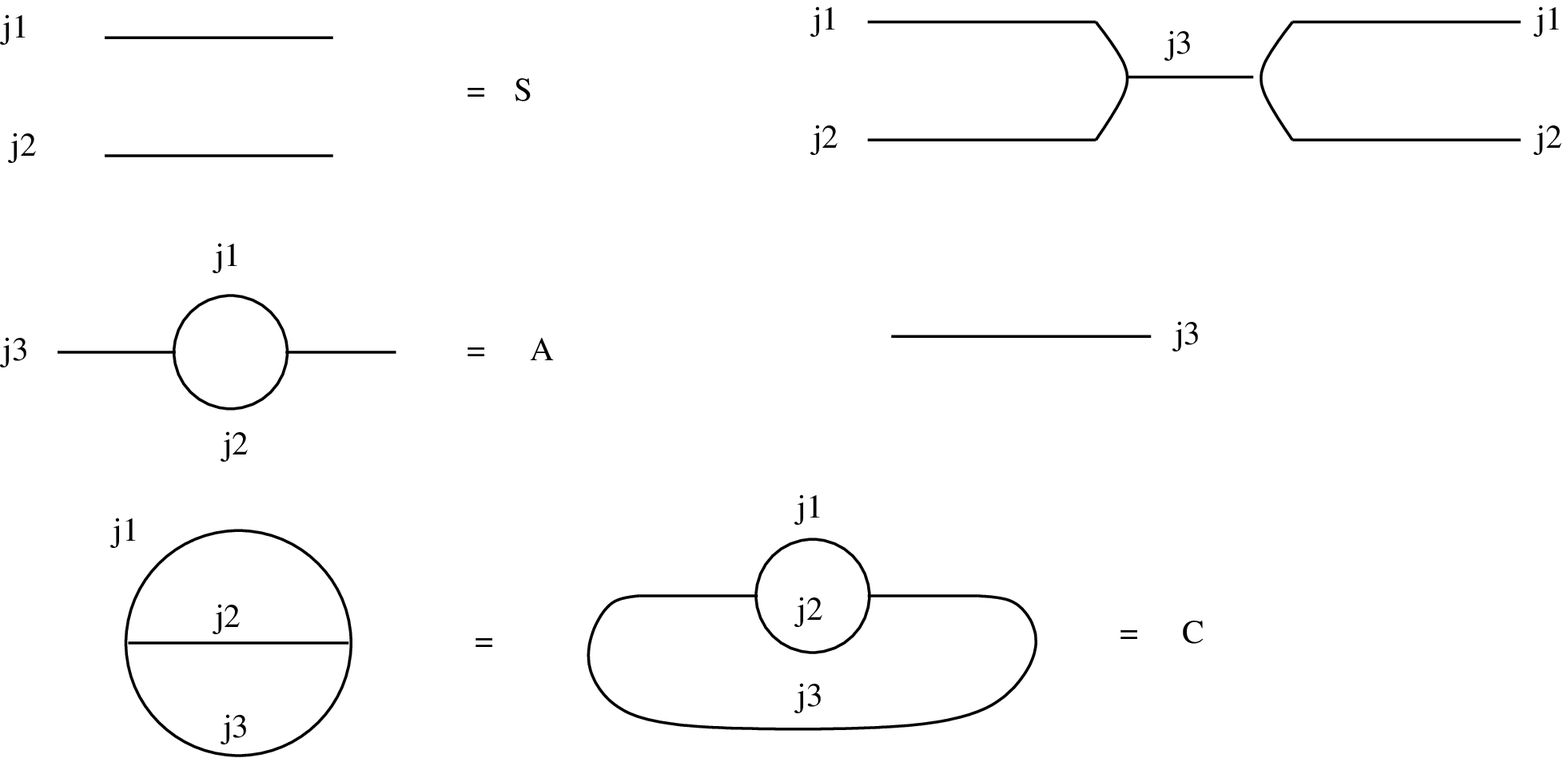}
\caption{Illustration of the normalization properties: the three relations are in fact equivalent
to the condition that the $\theta$-graph is normalized to one.}
\label{norm}
\end{figure}

In the sequel, we fix the normalization of $\iota(j_1,j_2;j_3)$ such that it satisfies the following fusion rule:
\be
 \sum_{j_3} [d_{j_3}] \, i(j_1,j_2;j_3) \cdot i(j_3;j_1,j_2) \; = \; \mathbb I_{j_1\otimes j_2}
\ee
where $\mathbb I_{j_1 \otimes j_2}$ is the identity map in the tensor product of the two representations $j_1$ and $j_2$
and $\iota(j_3;j_1,j_2):V_{j_3} \rightarrow V_{j_1} \otimes V_{j_2}$ is the adjoint of $\iota(j_1,j_2;j_3)$. This relation implies
that the so-called $\theta$-graph is normalized to one
\be
\text{Tr}_{j_1\otimes j_2} (\iota(j_1,j_2;j_3) \cdot \iota(j_3;j_1,j_2)) \; = \;
\text{Tr}_{j_3} (\iota(j_3;j_1,j_2) \cdot  \iota(j_1,j_2;j_3)) \; = \; Y(j_1,j_2,j_3) \;,
\ee
which is equivalent to
\be
\iota(j_3;j_1,j_2) \cdot  \iota(j_1,j_2;j_3) \; = \; \frac{1}{[d_{j_3}]} \, Y(j_1,j_2,j_3) \,\mathbb I_{j_3}\,.
\ee
We made used of the notation $Y(j_1,j_2,j_3) \in \{0,1\}$ which is one only when $(j_1,j_2,j_3)$ satisfy the triangular inequalities,
otherwise it vanishes. These identities are graphically represented in the figure (\ref{norm}).

\medskip

Of particular interest for the quantization of Chern-Simons theory is the fact that $U_q(su(2))$ is quasi-triangular and
therefore admits an universal R-matrix which is at the origin of the braiding properties associated to the quantum groups. Without
entering too much into the details, let us recall that
$R \in U_q(su(2))\otimes U_q(su(2))$ satisfies in particular the so-called quantum Yang-Baxter equation and other defining
properties that one can find in \cite{CP} for example.

\begin{figure}[h]
\psfrag{j1}{$j_1$}
\psfrag{j2}{$j_2$}
\psfrag{R}{$R$}
\psfrag{R-1}{$R^{-1}$}
\centering
\includegraphics[scale=0.5]{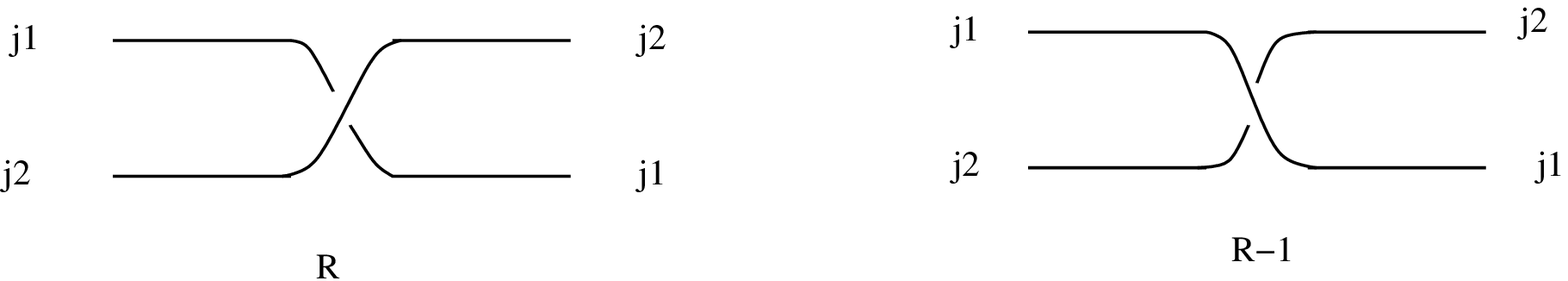}
\caption{Pictorial representation of the R-matrix and its inverse $R^{-1}$. Both R-matrices are evaluated in $j_1\otimes j_2$.}
\label{braiding}
\end{figure}

The evaluation of the R-matrix in the tensor product of representations $j_1 \otimes j_2$
is denoted $R_{j_1j_2}=(\pi_{j_1}\otimes \pi_{j_2})(R)$ and defines a braiding operator from $V_{j_1}\otimes V_{j_2}$
to the opposite tensor product $V_{j_2} \otimes V_{j_1}$.
It is useful to represent the R-matrix as in the picture (\ref{braiding}):
if $R$ is represented by an under-crossing (the up-line undercrosses the down-line)
then its inverse $R^{-1}$ is represented by an over-crossing (the up-line overcrosses the down-line).
It is clear from this representation that the product of $R$ by its inverse is the identity because the braiding has been unknoted.

\subsubsection*{1.2 From Verlinde coefficients to the Hilbert space dimension}
Now, we have all the ingredients to construct the Verlinde coefficients. These coefficients appeared first \cite{Verlinde}
in the context of conformal field theory
and then it has been realized that they have a very simple algebraic interpretation in the context of quantum groups.
Here we will give only their algebraic definition
and some of their properties which are important in the calculation of the dimension of the Hilbert space $H_k(j_1,\cdots,j_\ell)$.

\medskip

\begin{figure}[ht]
\psfrag{equiv}{$\equiv$}
\centering
\includegraphics[scale=0.5]{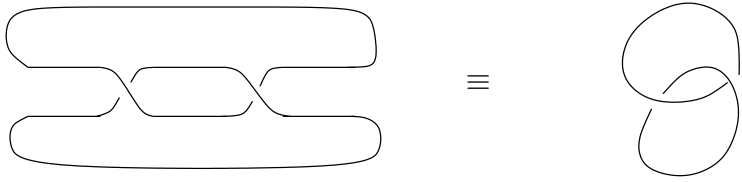}
\caption{Representation of the Hopf-link. The evaluation of the associated quantum spin-network colored with the representations
$j_1$ and $j_2$ gives the un-normalized Verlinde coefficient $\tilde{S}_{j_1j_2}$.}
\label{Hopflink}
\end{figure}

Given two unitary irreducible representations $j_1$ and $j_2$, one defines the Verlinde coefficient $S_{j_1j_2}={S}_{j_2j_1}$ as
the real number determined by the trace on $V_{j_1}\otimes V_{j_2}$ of the operator $R^2$ up to a normalization factor:
\be
{S}_{j_1j_2} \; = \; {\cal Z} \, {\text{Tr} \, (R_{j_2 j_1} R_{j_1j_2})}
\ee
where
\be
{\cal Z}=\sqrt{\frac{2}{k+2}} \,{\sin(\frac{\pi}{k+2})}
\ee
is in fact the partition function of the $SU(2)$ Chern-Simons theory on the 3-sphere $S^3$.
It will be useful in the sequel to use the ``un-normalized" Verlinde coefficient $\widetilde{S}_{j_1j_2}=\text{Tr} \, (R_{j_2 j_1} R_{j_1j_2})$
and the choice of the normalization factor will appear clear soon. Note that $\widetilde{S}_{j_1j_2}$ is the evaluation on the Hopf-link
embedded into the 3-sphere. The Hopf-link is represented in the figure (\ref{Hopflink}).

The explicit expression of the R-matrix implies that
\be\label{explicitVerlinde}
\widetilde{S}_{j_1j_2} \; = \; [d_{j_1}d_{j_2}] \; = \; \frac{\sin(\frac{\pi d_{j_1} d_{j_2}}{k+2})}{\sin(\frac{\pi}{k+2})}\;.
\ee

\begin{figure}[ht]
\psfrag{j1}{$j_1$}
\psfrag{j2}{$j_2$}
\psfrag{j3}{$j_3$}
\psfrag{l}{$\ell$}
\psfrag{=}{$=$}
\psfrag{A}{$[d_{j_3}]^{-1}$}
\psfrag{B}{$\sum_\ell Y(j_1,j_2,\ell)$}
\centering
\includegraphics[scale=0.5]{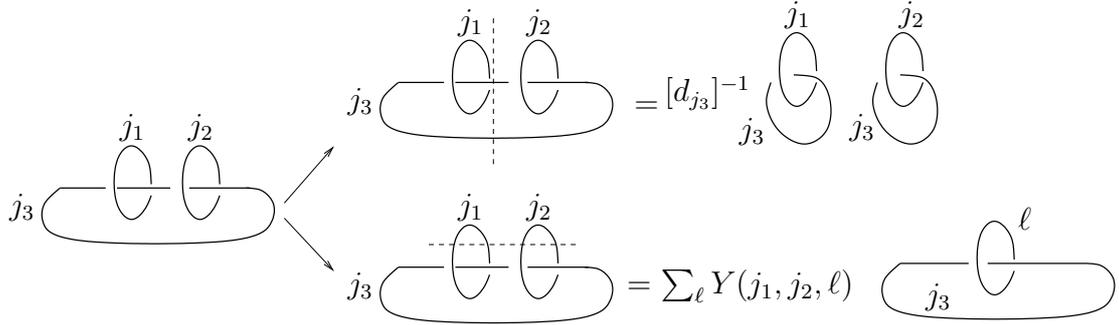}
\caption{Pictorial proof of the fusion relation. We start with the graph on the left. The two arrows are identities: the first one is obtained
applying the decomposition of the identity ``along the vertical dashed line''; the second one is obtained
applying the decomposition of the identity ``along the horizontal dashed line''. Both lead to equivalent expressions and the equality between the
evaluations of the graphs on the right is exactly the fusion relation. We made used of the identities represented in the picture (\ref{norm}).}
\label{proof}
\end{figure}

These coefficients satisfy many interesting properties which are important to compute the dimension of the physical Hilbert space
$H_k(j_1,\cdots,j_\ell)$ presented above. The properties we will need are given below:
\ba
&&\text{The normalization relation:} \;\;\; \sum_{j_3} \widetilde{S}_{j_1j_3} \widetilde{S}_{j_3j_2} \; =
\; \frac{\delta_{j_1j_2}}{{\cal Z}^2}; \label{normalization}\\
&&\text{The fusion relation:} \;\;\; \widetilde{S}_{j_1j_3} \widetilde{S}_{j_2j_3} \; = \; [d_{j_3}] \sum_{\ell} Y(j_1,j_2,\ell) \; \widetilde{S}_{j_3\ell};  \\
&&\text{The recursive relation:} \prod_{i=1}^{n-1}\widetilde{S}_{j_ij_n}   =  [d_{j_n}]^{n-2} \!\!\sum_{\ell_1,\cdots,\ell_n} \!\!\delta_{\ell_1,0}
\prod_{i=1}^{n-1} Y(j_i,\ell_i,\ell_{i+1})  \widetilde{S}_{\ell_n j_n}.\label{recursive}
\ea
The definition of the normalized Verlinde coefficient becomes clear from the normalization relation.
The recursive relation is a generalization of the fusion relation to any number of unitary irreducible representations
${\bf j}=(j_1,\cdots,j_p)$. The fusion and recursive relation are really easy to prove using
the graphical representations of the Verlinde coefficients. The proof of the fusion relation is given in the picture
(\ref{proof}); the proof of the recursive relation is done along exactly the same lines.

\medskip

Verlinde coefficients and their properties are particularly interesting to obtain useful formulae for the dimension of
the Hilbert space $H_k(j_1,\cdots,j_p)$.
Indeed, the dimension $N_k({\bf j})=\text{dim}(H_k(j_1,\cdots,j_p))$ is
\be
N_k({\bf j}) \; = \; \sum_{\ell_1,\cdots,\ell_p} \delta_{\ell_1,0} \delta_{\ell_{p+1},0} \prod_{i=1}^p Y(\ell_i,j_i,\ell_{i+1})
\ee
and can be expressed in terms of Verlinde coefficients using
the recursive relation (\ref{recursive}) combined with the normalization relation (\ref{normalization}). Some trivial
calculations lead to the expression:
\ba
N_k({\bf j}) \; = \; \frac{2}{k+2} \sin^2(\frac{\pi}{k+2}) \sum_{\ell}[d_\ell]^{2-p} \prod_{i=1}^p \widetilde{S}_{j_i\ell}
\ea
which reduces, after using the explicit formula of Verlinde coefficients (\ref{explicitVerlinde}), to the following well-known formula:
\be\label{dimension}
N_k({\bf j}) \; = \; \frac{2}{k+2} \sum_\ell (\sin(\frac{\pi d_\ell}{k+2}))^{2-p}\prod_{i=1}^p \sin(\frac{\pi d_\ell d_{j_i}}{k+2}) \;.
\ee

\subsection*{2. Equivalent formulae for the Hilbert space dimension}
The expression (\ref{dimension}) for the dimension of the $SU(2)$ Chern-Simons Hilbert space is not very useful to compute the entropy of
a Black Hole. We propose here to give equivalent more interesting formulae.

\subsubsection*{2.1. Chern-Simons Hilbert space and random walk}
The fact that the coefficient $N_k({\bf j})$ are closely related to random walks have been noted and investigated in \cite{livi}
in the classical case, namely when $k$ becomes infinite. Here we show that even in the quantum case (i.e. for a finite $k$)
this link between Chern-Simons and random walk still exists and appears very interesting for the calculation of the entropy.
Some of our formulae have been derived in \cite{kaulma} where however only the classical case (infinite $k$) has been studied at the end.
For obvious reasons of notations, we will consider $\widetilde{N}_k({\bf d})\equiv N_{k-2}({\bf j})$ in the sequel.
We will also make use of the notations  $d_i=d_{j_i}=2j_i+1$ and ${\bf d}=(d_1,\cdots,d_p)$.

\medskip

This Section is devoted to propose a random walk interpretation of the dimension $\widetilde{N}_k({\bf d})$.
 For that purpose, we proceed in four steps.

\subsubsection*{a) $\widetilde{N}_k({\bf d})$ as a difference of $B_k({\bf d})$-type functions}
To do so, we first use the identities $\sin^2\theta= 1-\cos^2\theta$ and $\cos \theta =\sin(2\theta)/(2\sin^2\theta)$
in the formula
\be
\widetilde{N}_k({\bf d}) \; = \; \frac{2}{k} \sum_\ell (\sin(\frac{\pi d_\ell}{k}))^{2}\prod_{i=1}^p
\frac{\sin(\frac{\pi d_\ell d_{j_i}}{k})}{\sin(\frac{\pi d_\ell}{k})}
\ee
to write it as the difference
\be\label{NfunctionofB}
\widetilde{N}_k({\bf d}) \; = \; B_k({\bf d}) - \frac{1}{4} B_k({\bf d_+})
\ee
where the function $B_k$ which depends on a family of representations ${\bf d}=(d_1,\cdots,d_p)$ or
${\bf d_+}=(d_1,\cdots,d_p,2,2)$ reads
\be\label{definitionofB}
B_k({\bf d}) \; = \; \frac{2}{k} \sum_{d=0}^{k-1} \prod_{\ell} \frac{\sin(\frac{\pi d}{k}d_\ell)}{\sin(\frac{\pi d}{k})}\,.
\ee
The product runs over $\ell \in [0,p]$ or $\ell \in [0,p+2]$ depending whether we are considering $B_k({\bf d})$
or $B_k({\bf d_+})$.
Note that ${\bf d_+}$ is the union of the family of dimensions $\bf d$ with two more equal elements corresponding to the dimension
of the fundamental representation $d_{1/2}=2$.

\subsubsection*{b) Combinatorial expression of $B_k({\bf d})$}
Now  we  concentrate on the function $B_k({\bf d})$. In particular, we want to exhibit the fact, as in the classical case,
that $B_k({\bf d})$ admits a random walks interpretation. To show this is indeed the case,
we replace each term of the product in (\ref{definitionofB}) by the following expression:
\be
\frac{\sin(\frac{\pi d}{k}d_\ell)}{\sin(\frac{\pi d}{k})} = e^{i\frac{\pi d}{k}(d_\ell -1)} \sum_{n_\ell=0}^{d_\ell -1} e^{-2i\frac{\pi d}{k}n_\ell}\;.
\ee
As a result, the function (\ref{definitionofB}) can be rexpressed as follows:
\ba
B_k({\bf d})  =  \frac{2}{k} \sum_{d=0}^{k-1} \prod_{\ell =1}^p \sum_{n_\ell =0}^{d_\ell -1} e^{i\frac{\pi d}{k}(d_\ell -1 -2n_\ell)}
 =  \frac{2}{k} \sum_{d=0}^{k-1} \sum_{\{n_1,\cdots,n_p\}} e^{i\frac{\pi d}{k}(\Delta_p -2N)}
\ea
where we introduced the notations $\Delta_p=\sum_{\ell =1}^p(d_\ell -1)$ and $N=\sum_{\ell =1}^p n_\ell$. Note that
the second sum runs over families of integers $\{n_1,\cdots,n_p\}$ such that each component $n_i\in [0,d_i-1]$.
Permuting the two sums and summing over the variable $d$ lead to:
\ba\label{Bk}
B_k({\bf d}) \; = \; \frac{2}{k} \sum_{\{n_1,\cdots,n_p\}} \frac{1 - e^{i\pi (\Delta_p -2N)}}{1 - e^{i\frac{\pi}{k}(\Delta_p - 2N)}} =
\frac{2}{k} \sum_{\{n_1,\cdots,n_p\}} \frac{1 - e^{i\pi \Delta_p}}{1 - e^{i\frac{\pi}{k}(\Delta_p - 2N)}}
\ea
To go further into the calculation, we distinguish the case where $\Delta_\ell$ is odd from the case where $\Delta_\ell$ is even.

\subsubsection*{c) $\Delta_\ell$ odd implies that $\widetilde{N}_k({\bf d})=0$}
The first case, $\Delta_\ell$ odd, is simpler. Indeed, in that case, $1-e^{i\pi\Delta_p}=2$ and then
the function $B_k({\bf d})$ reduces to the form:
\be
B_k({\bf d}) \; = \; \frac{4}{k}\sum_{\{n_1,\cdots,n_p\}} \frac{1}{1 - e^{i\frac{\pi}{k}(\Delta_p - 2N)}} \,.
\ee
From the beginning, we know that $B_k({\bf d})$ is a real-valued function and therefore it equals its real part, i.e.
$B_k({\bf d})={\cal R}_e(B_k({\bf d}))$ where ${\cal R}_e(z)$ denotes
the real part of $z \in \mathbb C$. Moreover, for any value of $\theta$ (different from $0[2\pi]$), the following equality holds:
\be
\frac{1}{1-e^{i\theta}} = \frac{1}{2} + \frac{i}{2} \text{cotan} \frac{\theta}{2}\;.
\ee
As a consequence, the expression of the function $B_k({\bf d})$ simplifies drastically and reduces to:
\be
B_k({\bf d}) =\frac{4}{k} \frac{1}{2} \sum_{\{n_1,\cdots,n_p\}} 1 \; = \; \frac{2}{k} \prod_{\ell =1}^p d_\ell \;.
\ee
Therefore, $B_k({\bf d_+})=4B_k({\bf d})$ and then the dimension of the Hilbert space (\ref{NfunctionofB}) vanishes
in that case. The meaning of this result is simple:
there is no invariant tensor in the tensor product $\otimes_\ell j_\ell$ when $\Delta_p =\sum_\ell (d_\ell -1)$ is odd.
As an example, let us consider the case where all the spins equal $1/2$: $\Delta_p=p$ odd means that there is an odd number
of spins; as expected there is no trivial representation in the tensor product of an odd number of $1/2$ representations.

\subsubsection*{d) $\Delta_\ell$ even: random-walk interpretation of $\widetilde{N}_k({\bf d})=0$}
The second case, $\Delta_p$ even, is far more interesting. In that situation, we would naively say that $B_k({\bf d})$ vanishes because
all the terms in the numerator of the formula (\ref{Bk}) are $1-e^{i\pi \Delta_p}=0$. But a more careful analysis shows that the
denominator can also lead to a singularity. As a result, the non-vanishing contributions to the sum (\ref{Bk}) are those where both
the numerator and denominator vanish. For this to happen, there must exist $s\in \mathbb Z$ such that $\Delta_p-2N=2sk$.
Therefore, we have:
\be\label{random}
B_k({\bf d})=2 \sum_{\{n_1,\cdots,n_p\}}\delta_{\Delta_p -2N[2k]}
\ee
where $\delta_{n[2k]}$ takes the value one if there exists an integer $s$ such that $n=2ks$, otherwise it is null.
Here comes the random walk interpretation of the dimension of the Chern-Simons Hilbert space. We proceed to a changing of
variables: instead of summing over non-negative integers $n_i \in [0,d_i-1]$, we sum over half-integers
$m_i = n_i - \frac{d_i-1}{2} \in [\frac{1-d_i}{2},\frac{1+d_i}{2}]$ (with $m_{i+1}=m_i+1$). Then, the formula (\ref{random}) becomes
\be
B_k({\bf d})=2 \sum_{\{m_1,\cdots,m_p\}} \delta_{m_1+\cdots+m_p[k]}\,.
\ee
A similar formula has been found in \cite{kaulma} and its classical counterpart has been established and studied in \cite{livi}.
As a result, the $B_k$ function appears to be the number of ways to start from the origin $0$ at the $\mathbb Z$ axe and come back at a point
$0[k]$  after $p$ steps $m_i$, each  step being bounded as follows $m_i \in [\frac{1-d_i}{2},\frac{1+d_i}{2}]$. These functions have been deeply and
precisely studied in the domain of random walks. And this analogy was used as a central tool in \cite{livi} to obtain asymptotics behavior
of some entropy. In order to be a bit more explicit, we introduce the variable $r=[\Delta_p/(2k)]$ where $[x]$ is the floor function and then:
\be
B_k({\bf d})=2 \sum_{\{m_1,\cdots,m_p\}} \sum_{q=-r}^r \delta_{m_1+\cdots+m_p - qk} \;.
\ee
Let us recall now that we are interested in the number of states $\widetilde{N}_k$ and not in the function $B_k$ itself.
Using previous formulae, we have for $\widetilde{N}_k$ the expression:
\ba
\widetilde{N}_k({\bf d}) = 2 \sum_{\{m_1,\cdots,m_p\}} \left( \sum_{q=-r}^r \delta_{m_1+\cdots+m_p - qk} -
\frac{1}{4}\sum_{a,b \in \{-\frac{1}{2},\frac{1}{2}\}} \sum_{q=-s}^s \delta_{m_1+\cdots+m_p+a+b - qk} \right) \nonumber
\ea
where $s=[(\Delta_p+2)/(2k)]$. It is clear that $s$ belongs to the set $\{r,r+1\}$ and to avoid  complications
we assume that $r=s$. The case $s=r+1$ would introduce extra terms which are not important at all for what we want to do.
In that case, the previous formula simplifies and after summing over the variables $a$ and $b$ one obtains:
\be\label{randomN}
\widetilde{N}_k({\bf d})  = \sum_{\{m_1,\cdots,m_p\}} \sum_{q=-r}^r  ( \delta_{m_1+\cdots+m_p - qk} - \frac{1}{2}\delta_{m_1+\cdots+m_p - qk +1}
-\frac{1}{2}\delta_{m_1+\cdots+m_p - qk -1})\;.
\ee
This expression generalizes the one obtained in the classical case (which corresponds in fact to $r=0$ in our notations).
It is useful to study the asymptotic behavior of the number of states and also to study the effect of a finite $k$.

\subsubsection*{2.2. Integral formula}
Very often, one identifies the number of states $\widetilde{N}_k$ to the dimension of the invariant tensors space
in the tensor product $\otimes_\ell j_\ell$
between representations of the classical group $SU(2)$. This is only true when the ratio $r=0$ which also coincides with the classical limit $k$ goes
to infinity. In that case, $\widetilde{N}_k=\widetilde{N}_\infty$ is expressed as an integral over $SU(2)$ conjugacy classes, or equivalently
over an angle $\theta$.
We proposed to generalize this integral formula to the case where $r \neq 0$.

For that purpose, we start with the relation:
\ba
\delta_{m_1+\cdots+m_p + a} \; = \; \frac{1}{2\pi} \int_0^{2\pi}d\theta \; e^{i\theta(m_1+\cdots+m_p+a)}
\ea
defined for any integer $a \in \mathbb Z$. It easily leads to the relation:
\ba
\sum_{\{m_1,\cdots,m_p\}}\delta_{m_1+\cdots+m_p + a} \; = \; \frac{1}{2\pi} \int_0^{2\pi}d\theta
\cos(a\theta) \prod_{\ell=1}^p \frac{\sin(d_\ell\frac{\theta}{2})}{\sin \frac{\theta}{2}}\;.
\ea
where the sum runs over $m_\ell \in [\frac{1-d_\ell}{2},\frac{1+d_\ell}{2}]$.
Using this last identity and after some trivial calculations, one shows that the number of states is given by the integral:
\be
\widetilde{N}_k({\bf d}) \; = \; \frac{1}{\pi} \int_0^{2\pi} d\theta \; (\sum_{q=-r}^r \cos(\theta qk)) \sin^2(\frac{\theta}{2}) \prod_{\ell=1}^p \frac{\sin(d_\ell\frac{\theta}{2})}{\sin \frac{\theta}{2}}\;.
\ee
One more simplification occurs due to the trigonometric identity
\be\label{identitytrigo}
\sum_{q=-r}^r \cos(\theta qk) = 1 + 2 \sum_{q=1}^r\cos(qk\theta) = \frac{\sin((r+\frac{1}{2})k\theta)}{\sin \frac{k\theta}{2}} \;.
\ee
As a result, the number of states $\widetilde{N}_k$ takes the form:
\be\label{generalformula}
\widetilde{N}_k({\bf d}) \; = \; \frac{1}{\pi} \int_0^{2\pi} d\theta \;  \sin^2(\frac{\theta}{2})
 \frac{\sin((r+\frac{1}{2})k\theta)}{\sin \frac{k\theta}{2}}  \prod_{\ell=1}^p \frac{\sin(d_\ell\frac{\theta}{2})}{\sin \frac{\theta}{2}}\;.
\ee
This formula generalizes, as announced in the introduction of that section,
the classical one. We see, as expected, that $\widetilde{N}_k({\bf d})$ coincides with the classical formula when $r=0$,
i.e. when $\Delta_p < 2k$. This particular case  can be recovered from different arguments: if $\Delta_p <2k$, then each representations in the tensor
product $\otimes_\ell j_\ell$ has a spin $s <k/2$ and therefore one never sees the effect of the cut-off $k$.
When the condition $r=0$ is not satisfied the integral formula defining the number of states differs from the classical one by a different integration measure on the $SU(2)$ conjugacy class. The presence of this new measure might have an effect on the black hole entropy.

\subsubsection*{2.3. An example: all spins equal $1/2$}
To get an intuition of previous formulae, we consider a particular example:
we assume that $d_\ell=2$ for all punctures $\ell \in \{1,\cdots,p\}$, i.e. all the spins equal $1/2$.
Furthermore, we assume that $\Delta_p=p$ is even.

\subsubsection*{a) Classical case: $r=0$ \cite{livi,kaulma}.}
To start with, we also consider first the case $\Delta_p <2k$, i.e. $r=0$ in the previous
notations. This case has been studied deeply in \cite{livi}.

From the random walk expression of the number of states (\ref{randomN}),
one obtains that $\widetilde{N}_k(2,\cdots,2) \equiv {N}_{2}(p)$ is given by:
\be\label{classical1/2}
{N}_{2}(p) \; = \; \sum_{\{m_1,\cdots,m_p\}}(\delta_{m_1+\cdots+m_p} - \delta_{m_1+\cdots+m_p+1}) \; = \;
\binom{p}{p/2} - \binom{p}{p/2-1} \;.
\ee
We have omitted to mention $k$ because $\widetilde{N}_k$ does not in fact depend on $k$ when $r=0$.
The last equality is a result of a trivial combinatorial analysis: given an integer $a \leq p/2$, the number of ways to have
$n_1+\cdots+n_p=a$ where $n_i\in \{-1/2,1/2\}$ is given by the number
of ways to choose $(p/2+a)$ elements from a set of $p$ elements which is precisely given by the binomial coefficient $\binom{p}{p/2+a}$.
From the expression of the binomial coefficients in terms of factorials, one ends up with the following formula:
\be
{N}_{2}(p) \; = \; \frac{2}{p+2}\frac{p!}{((p/2)!)^2}
\ee
We obtain an exact combinatoric formula for the number of states in that particular case. The asymptotic of ${N}_2(p)$
is therefore straightforward to obtain from the stirling formula which states that:
\be\label{stirling}
p! \; \sim \; \sqrt{2\pi p} (p/e)^p \;\;\;\; \text{for large values of $p$.}
\ee
Using this very well-known result, one shows the following asymptotic behavior:
\be\label{F2}
{N}_2(p) \; \sim \; \sqrt{\frac{8}{\pi}}p^{-3/2} 2^p \;.
\ee
This formula coincides with the one found in \cite{kaulma} from different arguments. In particular, we recover the same leading order and
the same sub-leading corrections to the ``entropy":
\be
S_{1/2}(p) \; \equiv \; \log{N}_2(p)\; = \; 2^p \; - \; \frac{3}{2} \log p \; + \; {\cal O}(1).
\ee

\subsubsection*{b) Quantum corrections: $r>0$.}
Let us now relax the condition that $\Delta_p<2k$, i.e. $r=[p/(2k)]$ is now a non-zero integer. In that case, it is
a bit more involved to obtain a combinatoric expression for the number of states but, using similar arguments as previously,
one can show that:
\be
{N}_2(p) \; = \; \sum_{q=-r}^r  \binom{p}{p/2-qk} -\frac{1}{2}\binom{p}{p/2-qk-1} - \frac{1}{2} \binom{p}{p/2-qk+1}\;.
\ee
Note that we still omit to mention explicitly the dependence in $k$ even if now $N_2(p)$ does depend on $k$.
To go further, we separate the $q=0$ contribution from the others in the sum and, using trivial symmetries properties of binomial coefficients, we get:
\ba
{N}_2(p) & = & \binom{p}{p/2} - \binom{p}{p/2-1} \nonumber \\
 & & +\sum_{q=1}^r  2\binom{p}{p/2-qk} -\binom{p}{p/2-qk-1} - \binom{p}{p/2-qk+1}\,.
\ea
Thus, we obtain a correction to the previous classical case (\ref{classical1/2}) and each term (for a given value of $q \in [1,r]$)
in the remaining sum is a finite linear
combination of binomial coefficients which reduces to the following form after some calculations:
\be\label{remainingterms}
2\frac{1-p-2q^2k^2}{(p/2+1)^2 -q^2k^2} \binom{p}{p/2-qk}\;.
\ee
As a result, the asymptotic behavior of each term in the previous sum is governed by the asymptotic behavior of the binomial coefficient
$\binom{p}{p/2-qk}$. We are interested in the asymptotic for a large value of $p$ but also a large value of $k$ such that these two numbers have
the same scaling namely $p/k$ remains constant. Indeed, in the black hole context, both $p$ and $k$ are proportional to the area of the horizon
which tends to infinity (in unit of Planck area). Thus, the leading term in the asymptotic expansion of (\ref{remainingterms}) is given by
the leading term of the expansion of binomial coefficient of the form $\binom{p}{\alpha p}$ for large values of $p$ where $\alpha \in [0,1/2[$.
From Stirling formula (\ref{stirling}), one shows that
\be
\binom{p}{\alpha p} \sim \frac{1}{\sqrt{2\pi \alpha(1-\alpha)p}}g(\alpha)^{-p} \;\;\; \text{with} \;\;\; g(\alpha)=
\alpha^\alpha(1-\alpha)^{1-\alpha}\;.
\ee
It is straightforward to show that the function $g$ satisfies the bound $g(\alpha) > 1/2$ and therefore the previous binomial
coefficient grows like $g(\alpha)^{-p} < 2^p$. As a consequence, the ``classical term" dominates the asymptotics in the sense that:
\be
{\binom{p}{\alpha p}}/{\binom{p}{p/2}} \; \sim \; (2g(\alpha))^{-p} \; \rightarrow \; 0
\ee
for $\alpha<1/2$. This result shows that the quantum corrections (due to the finiteness of $k$) do not affect the asymptotic
expansion of ${N}_2(p)$ neither at the leading neither at the subleading order. 

\subsection*{3. Entropy of the $SU(2)$ Black Hole}
Here we adapt the techniques used in \cite{counting3} and firstly introduced in \cite{counting1} to compute the entropy
of a black hole. We are not going into the mathematical details of these techniques which has been very well exposed
in \cite{counting3} and which  are in fact very well-known in the domain of probabilities and used
to understand some properties of random walks.
We propose to reproduce these results in a more ``intuitive" or physical way: we will omit many mathematical details
which appear in a first time non necessary. In particular, we show that it is not necessary to go to complex analysis and number theory
to get the asymptotic behavior of the entropy.

Before going to the details of the entropy calculation, let us briefly recall how black holes are described
in loop quantum gravity and how we compute the entropy.
In the context of LQG, a local definition of a black hole is introduced through the concept of {\em isolated horizons} (IH),  regarded as a sector of the phase-space of GR containing a horizon in equilibrium with the external matter and gravitational degrees of freedom. This local definition is used for the black-hole entropy calculation since the quantization of such a system allows to define a Hilbert space which is the tensor product of a boundary and a bulk terms.
The entropy of the IH is then computed by the formula $S={\rm tr}(\rho_{\va IH}\log\rho_{\va IH})$, where the density matrix
$\rho_{\va IH}$ is obtained by tracing over the bulk d.o.f., while restricting to horizon states that are compatible
with the macroscopic area parameter $a$. Assuming that there exists at least
one solution of the bulk constraints for every admissible state on the boundary, the entropy is given by
$S=\log(N(a))$ where $N(a)$ is the number of  horizon states. Since the theory on the horizon is associated to Chen-Simons theory with punctures, the entropy calculation problem boils down to the counting, in the large horizon area limit, of the dimension of the Hilbert space (\ref{Hilbert}), which, for a given configuration of punctures spins ${\bf d}=(d_1,\cdots,d_p)$, is expressed by the formula (\ref{generalformula}). 

\subsubsection*{3.1. The Laplace transform method: basic idea}
The Laplace transform method allows in certain cases to obtain the asymptotic behavior of a function
$F(p)$ for  large values of $p$.
For simplicity, we assume that the function $F$ is defined for $p$ integers
but the method applies in the case where $F$ is a function of a real number $x$.
The idea consists first in considering the Laplace transform $\widetilde{F}(s)$ defined a priori for $s \geq 0$ by:
\be\label{Laplace}
\widetilde{F}(s) \; = \; \sum_{p=0}^\infty e^{-ps} F(p) \;.
\ee
The Laplace transform appears as a series and therefore might be not defined at all or it might be defined for some values of the
real positive variable $s$ only. To understand if the series is convergent or divergent, one looks
at the asymptotic behavior of $F(p)$ at large $p$. To be more concrete, let us propose some examples.
\begin{enumerate}
\item If $F(p)\sim p^\alpha$ for some real number $\alpha$, then the series (\ref{Laplace}) is convergent for any values of $s$.
\item If $F(p)\sim e^{\alpha p^2}$ for some positive real number $\alpha$, then the series (\ref{Laplace}) is divergent for any values
of $s$ and then the Laplace transform is never defined. On the contrary, if $\alpha$ is negative, then the series in convergent
and the Laplace transform is well defined for all values of $s$.
\item If $F(p) \sim e^{\alpha p}$ for some positive real number $\alpha$, then the series is convergent for $s>\alpha$. In the case
where $s<\alpha$, the series diverges and therefore is ill-defined. The case $s=\alpha$ is critical: the convergence properties of $\widetilde{F}$
in that regime depends on the subleading behavior of $F(p)$.
\end{enumerate}
We are particularly interested in the last case because in the context of black hole entropy the number $N(a)$ of microstates corresponding
to a given macroscopic area $a$ is exponential in $a$, namely $N(a) \sim e^{\alpha a}$ and all the problem is to find the coefficient $\alpha$.
Let us now come back to the general discussion.
If we find $\alpha>0$ such  that $\widetilde{F}(s)$ is defined for $s>\alpha$ and undefined for $s<\alpha$, then we conclude that
$F(p)\sim e^{\alpha p}$ for large $p$. This is more a physical argument than a rigorous proof because we assume the asymptotic
behavior of $F(p)$. However, it is also possible to prove rigorously the asymptotic behavior as it was done in \cite{counting1}
and \cite{counting3} where the reader can find the mathematical details.
In the language of probabilities, $s_c=\alpha$ is called the critical exponent of $F(p)$.

Before going further, let us recall that, in the particular case where $F(p)\sim e^{\alpha p}$, it is possible to invert
the Laplace transform and to recover to function $F(p)$ from $\widetilde{F}(s)$ according to:
\be
F(p) = \frac{e^{s_0 p}}{2\pi}\int_0^{2\pi} dx \; \widetilde{F}(s_0+ ix) e^{ixp}
\ee
where $s_0>\alpha$ is a real number.

\medskip

In fact, it is possible to extend this technique to obtain also the subleading terms in the asymptotics expansion of $F(p)$.
To understand this point, we assume that $F$ behaves as follows $F(p)\sim e^{\alpha p}Q(p)$ where $Q(p)$ is an algebraic function
whose dominant term at large $p$ is $Q(p)\sim p^{\beta}$ where $\beta$ is a real number. We generalize the Laplace transform
to the following function of the two real variables $s$ and $t$:
\be\label{generalLaplace}
\widetilde{F}(s,t) = \sum_{p=1}^\infty e^{-ps} \, p^{-t} \, F(p)\;.
\ee
When $t=0$, this function coincides with the standard Laplace transform up to the constant $F(0)$. The point is to evaluate it at the critical value
$s_c=\alpha$. Indeed, $e^{-\alpha p}F(p) \sim p^{\beta}$ and therefore the convergence properties of the series $\widetilde{F}(\alpha,t)$ depends on
the asymptotic behavior of $p^{\beta-t}$: if $\beta-t>-1$, the series is divergent; if $\beta - t<-1$, the series is convergent. Therefore, we proceed
as we do to obtain $\alpha$, we define the critical value $t_c$ as the minimal value of $t$ such that $\widetilde{F}(\alpha,t)$ is well-defined.
Thus, the coefficient $\beta$ is fixed by $\beta=t_c-1$.

Of course, we can repeat this technique
to obtain recursively all the corrections to the leading term in the asymptotic expansion of $F(p)$. But, for this method to work,
we must know the form of the asymptotic behavior of the function $F(p)$. Once we know that the function $F(p)\sim e^{\alpha p} p^\beta$
at large $p$, our method allows to obtain the critical exponents $\alpha$ and $\beta$.

\subsubsection*{3.2. A simple application of the Laplace transform method}
Let us show that we can use this very simple technique to obtain the asymptotic behavior of the number of states $\widetilde{N}_k({\bf d})$
when all the spins are equal. To that aim, we introduce the notation $F_d(p)=\widetilde{N}_k({\bf d})$ with $j_1=\cdots=j_p=j$
and $d=2j+1$:
\be\label{pipo}
F_d(p) \; = \; \int_0^\pi d\theta \; \mu_k(\theta) \left(\frac{\sin (d\theta)}{\sin \theta}\right)^p
\ee
where $\mu_k(\theta)$ is a continuous non singular function obtain directly from (\ref{generalformula}):
\be
\mu_k(\theta) \; = \; \frac{2}{\pi}\, \sin^2\theta \, \frac{\sin((2r+1)k\theta)}{\sin(k\theta)}\;.
\ee
We omit to mention the dependence in $k$ of $\mu_k(\theta)$ and of $F_d(p)$ for clarity reasons.
Furthermore, we know, from random walks arguments, that the asymptotics of $F_d(p)$ is dominated by its classical part only;
in other words, we consider $r=0$ in the sequel. We will discuss the quantum corrections later.
Now, we assume that $F_d(p) \sim e^{\alpha p}$ for large values of $p$.
If the asymptotic behavior assumption is true, then  the Laplace transform
$\widetilde{F}_d(s) =  \sum_{p=0}^\infty e^{-sp}F(d,p)$
is well-defined for $s > \alpha$ and not-defined for $s<\alpha$. To obtain the critical exponent $\alpha$, we need to simplify the expression
of $\widetilde{F}_d(s)$. To do so, we exchange the sum defining $\widetilde{F}_d$ and the integral over $\theta$ in the definition of $F_d$
(\ref{pipo}). We obtain the following expression:
\be\label{integralF}
\widetilde{F}_d(s) \; = \; \frac{2}{\pi}\int_0^\pi d\theta \, \sin^2\theta \, \sum_{p=0}^\infty
\left(e^{-s} \frac{\sin (d\theta)}{\sin \theta}\right)^p
\ee
Then, assuming that $s$ is large enough, we perform the sum over $p$ and then:
\ba
\widetilde{F}_d(s)  = \frac{2}{\pi} \int_0^\pi d\theta \, \sin^2\theta \, Z(s,\theta) \;\;\; \text{with} \;\;\;
Z(s,\theta) =  \left( 1 - e^{-s}\frac{\sin (d\theta)}{\sin \theta}\right)^{-1} \nonumber \\
\ea
The next step consists in analyzing the structure of the singularities of $\widetilde{F}_d(s)$. It is clear that its singularities
come from the poles of the function $Z(s,\theta)$ viewed as a function of $\theta$.
We immediately see that $Z(s,\theta)$ admits a pole (viewed as a function of $\theta$)
if and only if $e^{-s} \geq d^{-1}$, i.e. $s \leq \log d$.  More precisely, we have the following:
\begin{enumerate}
\item When $s=\log d$, then  $Z(s,\theta)$
admits an unique pole which is $\theta_0=0$.
\item When $s < \log d$, then $Z(s,\theta)$ admits also at least one pole $\theta_0 \neq 0[\pi]$. At the vicinity of $\theta_0$, the
(inverse of the) function $Z$
behaves as follows:
$$Z^{-1}(s,\theta_0+\varepsilon) \simeq -\varepsilon(d\text{cotan}(d\theta_0) - \text{cotan}(\theta_0)).$$
As a consequence, the integral (\ref{integralF}) is divergent.
\end{enumerate}
As expected, $\widetilde{F}(d,s)$ is defined only for $s>\log d$. Therefore, $s_c=\log d$ is the
critical exponent and we have the asymptotic behavior:
\be
F_d(p) \; \sim \; d^p \;.
\ee
Before computing the sub-leading corrections, let us make some important remarks.

In the first remark, we come back to the exchange of the sum over $p$ and the integral over $\theta$ in the computation
of the Laplace transform (\ref{integralF}). This step can be justified in our case but this is not always the case.
More precisely, the exchange makes sense if the sum over $p$ is defined, namely if $\vert e^{-s}\sin(d\theta)/\sin\theta \vert <1$
for all $\theta$. This is exactly the condition we obtained to compute the critical exponent.

In the second and last remark, we come back to the ``quantum corrections" of $F_d(p)$. We know that the number of states is given by:
\ba
F_d(p) & = &  \frac{2}{\pi}\int_0^\pi d\theta \, \sin^2\theta \left(\frac{\sin (d\theta)}{\sin \theta}\right)^p \nonumber\\
& & +\frac{4}{\pi}\sum_{q=1}^r \int_0^\pi d\theta \, \sin^2\theta \, \cos(qk\theta)
\left(\frac{\sin (d\theta)}{\sin \theta}\right)^p \;.
\ea
The first term is the classical contribution and the rest are the quantum corrections. We want to compute the asymptotic behavior
of these corrections at the large $p$ limit, and we take,
at the same time, $k$ large as well with a fixed ratio $\rho=p/2k$.
As a consequence the number $r=[\rho]$ remains fixed in this limit. Therefore, the contribution of $F_d(p)$ corresponding to $q \neq 0$ reads:
\be
\frac{4}{\pi}\sum_{q=1}^r \int_0^\pi d\theta \, \sin^2\theta \, \cos(\frac{q\theta}{2\rho} p)
\left(\frac{\sin (d\theta)}{\sin \theta}\right)^p \;.
\ee
The calculation of the Laplace transform is more involved
in that situation due to the presence of a fast oscillating function in the integrand. A consequence is that the naive exchange of the infinite sum
over $p$ and the integral over $\theta$ is not justified.

\medskip

Now, we go further and compute the sub-leading terms. For that purpose, we concentrate on the classical contribution only (formally we take $t=0$)
and we define the general Laplace transform of $F_d(p)$ (\ref{generalLaplace}):
\be
\widetilde{F}_d(s,t) \; = \; \sum_{p=0}^\infty e^{-ps} \, p^{-t} \, F_d(p)\;.
\ee
Permuting the integral with the sum leads to the expression:
\ba\label{generalizedexample}
\widetilde{F}_d(s,t)  =  \int_0^\pi \mu_k(\theta) \sum_{p=0}^\infty p^{-t} \left( e^{-s} \frac{\sin(d\theta)}{\sin \theta}\right)^p
= \int_0^\pi \mu_k(\theta) \, \text{Li}_t\left(\frac{\sin (d\theta)}{d\sin \theta}\right)\,.
\ea
Indeed, we recognize the polylogarithm function $\text{Li}_t(z)$ defined for any couple of complex numbers $(t,z)$  such that
$\vert z \vert <1$ by the series:
\be
\text{Li}_t(z) \; = \; \sum_{p=1}^\infty p^{-t} \, z^p \;.
\ee
Following the general idea we described above, we first evaluate $\widetilde{F}_d(s,t)$ at the critical value $s_c=\log d$.
Then, we look for the critical value $t_c$ such that $\widetilde{F}_d(s_c,t)$ is well-defined for $t>t_c$ but not defined if $t<t_c$.
We know that $\widetilde{F}_d(s_c,t)$  might be not defined because of the singularity of the integrand
$\text{Li}_t\left(\frac{\sin (d\theta)}{d\sin \theta}\right)$ at $\theta=0$. To compute $t_c$, we analyze the behavior of the integrand
around $\theta=0$:
\begin{eqnarray*}
\text{Li}_t\left(\frac{\sin (d\theta)}{d\sin \theta}\right)  & \sim & \text{Li}_t(\theta^2\frac{1-d^2}{6}) \\
& \sim & \text{Li}_t(e^{\frac{1-d^2}{6}\theta^2}) \\
& \sim & \Gamma(1-t) \left(\frac{d^2-1}{6}\right)^{t-1} \theta^{2(t-1)}
\end{eqnarray*}
where $\Gamma(t)$ is the Gamma function. We used some asymptotic properties of the polylogarithm function. As a consequence, at the vicinity
of $\theta=0$, the  integrand of (\ref{generalizedexample}) behaves as:
\be
\mu_k(\theta) \, \Gamma(1-t) \left(\frac{d^2-1}{6}\right)^{t-1} \theta^{2(t-1)} \; \sim \; \frac{2}{\pi}  \, \Gamma(1-t) \left(\frac{d^2-1}{6}\right)^{t-1}\theta^{2t}
\ee
because $\mu_k(\theta)\simeq \frac{2}{\pi} \theta^2$ for $\theta$ small.
Therefore, the integral over $\theta$ is defined when $2t>-1$  and not defined when $2t<-1$; then the
critical value of $t$ is $t_c=-1/2$. As a consequence, we obtain the value
of the second critical exponent
$$\beta = -1 + t_c=-3/2$$
which is independent of the dimension $d$. Finally, we can establish that:
\be
F_d(p) \, \sim \, d^p p^{-3/2} \;\;\;\; \text{then}  \;\;\;\; \log F_d(p) = p \log d -\frac{3}{2} \log p + {\cal O}(1)\;.
\ee
In particular, we recover from another method the asymptotic behavior of $F_2(p)$ given in (\ref{F2}). This asymptotics has been obtained in
\cite{livi} from random walks arguments. The case $d=2$ has also been consider in \cite{kaulma}.

\subsubsection*{3.3. Asymptotic of the entropy}
We apply, the method illustrated above to compute the asymptotic behavior of a spherically symmetric and a distorted $SU(2)$ black hole
entropy in loop quantum gravity.

\subsubsection*{a) The spherically symmetric Black Hole}
The calculation of the entropy of the spherically symmetric $SU(2)$ black hole has been done precisely in \cite{counting3}
and has been investigated earlier in \cite{kaulma} when the level in infinite. In \cite{su22, su23}, we have shown that
the level $k$ and the area $a$ can be considered as independent variables. For that reason,
we are going to reformulate the results obtained in \cite{counting3} when $k$ is finite  focussing on physical arguments and avoiding the
number theory and the complex analysis aspects. These  aspects are not necessary to get the main ideas and the main results
if one assumes that the number of states grows exponentially with the area.

\medskip

\begin{figure}[ht]
\centering
\includegraphics[scale=1]{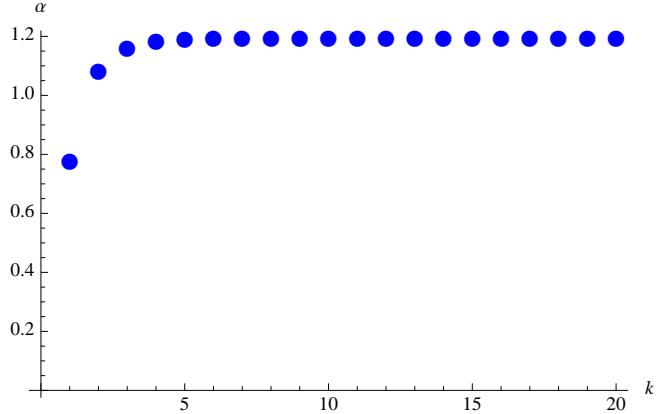}
\caption{Increasing critical values of the exponent $\alpha $ for different $k\in \mathbb N$. Already for $k\geq 4$ an asymptotic value is reached.}
\label{Fig:s-kSpheric}
\end{figure}

The entropy $S(a)=\log N(a)$ of a spherically symmetric black hole of macroscopic (adimensionalized) area $a$ is defined from
the number of states
\be\label{SSBH}
N(a) = \sum_{p=0}^\infty \sum^{k+1}_{d_1,\cdots,d_p} \delta(a-\frac{\sum_{\ell=1}^p\sqrt{(d_\ell-1)(d_\ell +1)}}{2}) \widetilde{N}_{k}({\bf j})\,.
\ee
The finiteness of the level $k$ appears in two different places: in the sums which run from $2$ to $k+1$ and in the expression
of $\widetilde{N}_{k}({\bf j})$. It is important to note again that we will consider $k$ and $a$ as independent variables
and we will study the entropy for large $a$ but finite $k$.
This consideration allows us to define the Laplace transform of $N(a)$
\be
\widetilde{N}(s)  =  \int_0^\infty da\, e^{-as}N(a) \,.
\ee
One difference with the previous Section is that now the variable $a$ is continuous.
After some calculations
and assuming that $s$ is large enough, we end up with the following expression:
\ba
\widetilde{N}(s) & = & \sum_{p=0}^\infty \sum^{k+1}_{d_1,\cdots,d_p} \int_{0}^\pi d\theta \, \mu_k(\theta)
\left( \prod_{\ell=1}^p \frac{\sin(d_\ell \theta)}{\sin \theta} e^{-\frac{s}{2} \sqrt{(d_\ell -1)(d_\ell +1)}}\right)\\
&=& \sum_{p=0}^\infty \int_0^\pi d\theta \,\mu_k(\theta) \left(\sum^{k+1}_{d=2}\frac{\sin d\theta}{\sin \theta} e^{-\frac{s}{2} \sqrt{(d-1)(d+1)}}\right)^p\\
&=& \int_0^\pi d\theta \, \mu_k(\theta) \left( 1 - \sum_{d=1}^k\frac{\sin (d+1) \theta}{\sin \theta} e^{-\frac{s}{2} \sqrt{d(d+2)}}\right)^{-1}
\ea
Again, we exchanged the sums over the variables $d_\ell$ and $p$ with the integral over $\theta$.
We proceed as in the previous section and conclude that the critical value of $s$ is the highest zero of the function
\be
1 - \sum_{d=1}^k \frac{\sin (d+1)\theta}{\sin \theta} e^{-\frac{s}{2} \sqrt{d(d+2)}}
\ee
which is reached for $\theta=0$. Therefore, the critical exponent $\alpha$ is the unique solution of the equation
\be\label{barberosum}
1 - \sum_{d=1}^k (d+1) \, e^{-\frac{\alpha}{2} \sqrt{d(d+2)}} \; = \; 0 \,.
\ee
With that definition, $\alpha$ depends on the level $k$.
For increasing values of the level $k$, the solutions $\alpha$ of the previous equation reach fast an asymptotic value as plotted in Figure \ref{Fig:s-kSpheric}.
The asymptotic value coincides as expected with the value $\alpha_{\infty}$ found in \cite{counting3} when $k\rightarrow \infty$.
Furthermore, we can estimate how the difference $\Delta \alpha= \alpha_\infty - \alpha$ decreases when $k$ increases.
$\Delta \alpha$ decreases exponentially with $k$ in the sense that it exits two real positive constants $A$ and $B$
such that
$$
\vert \alpha_k - \alpha_\infty \vert \; < \; A \, e^{-Bk} \,.
$$

\medskip

Next, we compute the subleading corrections. The difficulty of this problem is that we could not find a way
to put the generalized Laplace transform in a ``suitable" form as it was the case for the "toy-example"
we considered above. We will nonetheless circumvent this difficulty as follows.
First, we will evaluate $\widetilde{N}(s)$ at the vicinity of the critical value $\alpha$, i.e. $s=\alpha + \varepsilon$ for
$\varepsilon$ small, and then we will compute the generalized Laplace transform $\tilde{N}(\alpha,-t)$ at the critical value $\alpha$
as follows:
\be
\widetilde{N}(\alpha,-t) \; = \;  \frac{\partial^{t} \widetilde{N}(\alpha + \varepsilon)}{\partial \varepsilon^t}\vert_{\varepsilon=0}\;.
\nonumber
\ee
assuming that $t$ is an integer. Finally, we will see that it makes sense to extend the obtained formula to half-integers (and also
real numbers in fact) which will allows us to extract the sub-leading corrections to the entropy.

\medskip

Let us start, as announced, by the following calculation:
\ba\label{bla}
\widetilde{N}(\alpha + \varepsilon) \; \simeq \;
\int_0^\pi d\theta \, \mu_k(\theta) \left( 1 - \sum_{d=1}^{k} \frac{\sin (d+1)\theta}{\sin \theta} e^{-\frac{\alpha}{2} \sqrt{d(d+2)}}
(1-\frac{\varepsilon}{2} \sqrt{d(d+2)})\right)^{-1}
\ea
We know that the singularity of $\widetilde{N}(s+\varepsilon)$ when $\varepsilon$ goes to zero is due to the singularity of the integrand
$$f_\varepsilon(\theta) \; = \; \mu_k(\theta)  \left( 1 - \sum_{d=1}^{k} \frac{\sin (d+1)\theta}{\sin \theta} e^{-\frac{\alpha}{2} \sqrt{d(d+2)}}
(1-\frac{\varepsilon}{2} \sqrt{d(d+2)})\right)^{-1}$$
in (\ref{bla}) when $\theta$ goes to zero. Therefore, we concentrate on the behavior of $f_\varepsilon(\theta)$
at the vicinity of $\theta=0$:
\begin{eqnarray*}
f_\varepsilon(\theta)  \; \simeq \; (2r+1) \frac{2\theta^2}{\pi} \left( \sum_{d=1}^{k} (\varepsilon (d+1)\sqrt{d(d+2)} + \frac{\theta^2}{6}d(d+1)(d+2)) e^{-\frac{\alpha}{2}\sqrt{d(d+2)}}\right)^{-1}
\end{eqnarray*}
Now, we start from this expression to study the singularities of the generalized Laplace transform $\widetilde{N}(\alpha,t)$
evaluated at the critical value. Indeed, when $t$ is a positive integer, we can compute:
\be\label{generalizedLaplace}
\widetilde{N}(\alpha,-t) \; = \;  \frac{\partial^{t} \widetilde{N}(\alpha + \varepsilon)}{\partial \varepsilon^t}\vert_{\varepsilon=0}\;
\ee
The last quantity is expressed as an integral over the variable $\theta$ whose eventual singularity is due to the behavior of the integrand around
$\theta=0$ given by:
\ba
\frac{\partial^{t} f_\varepsilon(\theta) \vert_{\varepsilon=0}}{\partial \varepsilon^t} & \simeq &
(-1)^t t!\frac{2\theta^2}{\pi} \frac{ \left( \sum_{d=1}^k (d+1)\sqrt{d(d+2)}e^{-\frac{\alpha}{2} \sqrt{d(d+2)}}\right)^t}{\left( \frac{\theta^2}{6} \sum_{d=1}^k d(d+1)(d+2) e^{-\frac{\alpha}{2} \sqrt{d(d+2)}}\right)^{t+1}} \nonumber\\
& \sim &  \theta^2 \, \frac{1}{\theta^{2(t+1)}} = \theta^{-2t}\;.\nonumber
\ea
We assume that the behavior of the integrand of $\widetilde{N}(\alpha,-t)$ remains the same even if $t$ is any real number.
As a consequence, $\widetilde{N}(\alpha,t)$ is singular when $2t > 1$ and the critical value of $t$ is $t_c=1/2$.
Then, the critical exponent $\beta=-t_c-1=-3/2$ and we recover the asymptotic expansion \cite{counting3}:
\be
N(a) \; \sim \; e^{\alpha a} \, a^{-3/2}   \;\;\;\; \text{for large $a$},
\ee
with $\alpha$ given as in Figure \ref{Fig:s-kSpheric}.
Again, the finiteness of $k$ does not modify the sub-leading corrections when $k$ is large.

\medskip

Let us finish this section by two remarks. The first one concerns the effect of the finiteness of the level $k$ in the entropy.
As we have just seen, $k$ does  modify  the leading but does not modify the subleading corrections of the entropy. In that sense,
the logarithmic corrections seems to be universal and independent of the Immirzi parameter even in the $SU(2)$ spherically symmetric
black hole.
The second one concerns the techniques
we used: our calculations have to be viewed more as ``physical" arguments than rigorous proofs
of the asymptotic behavior of the entropy.
The nice point is that we can recover the ``right'' results very easily without entering too much into technical aspects.

\subsubsection*{b) The distorted Black Hole}
In the distorted case \cite{su23}, the black hole is described  in terms of two commuting Chern-Simons theories
associated to the same level $k$. As in the symmetric case, the area $a$ of the black hole and the level $k$ are
considered as independent variables. At the fundamental level, the description of the distorted black hole in terms
of microstates is rather different from the description of the symmetric black hole. Indeed, each puncture colored
with a $SU(2)$ representation $j$ coming from the bulk decomposes into two $SU(2)$ representations $j^+$ and
$j^-$ when it crosses the black hole. A macroscopic state is therefore characterized by the number $p$ of punctures and a
family $(j_\ell,j^+_\ell,j_-^\ell)_\ell$ ($\ell \in [1,p]$) of $3p$ representations of $SU(2)$ such that
$(j_\ell,j^+_\ell,j^-_\ell)$ satisfy the triangular inequality for each $\ell$ and an additional constraint.  

To describe this additional constraint, we associate canonically the $SU(2)$ generators $J_{+}^i$, $J_{-}^i$, and $J^i$
to each puncture: the Casimir of these operators $J_{\pm}^2$ and $J^2$ are fixed by the representations 
$j_\pm$ and $j$ in the standard way. 
The constraint reads 
\be\label{EPRL}
~C^i(p)= J_-^i-J_+^i-\alpha(J_+^i+J_-^i)=0.
\ee
with 
 \be\label{alfa}
\alpha\equiv\frac{ J_{+}^2-J_{-}^2}{J^2},
\ee
Now $C^i$ and $D^i=J^i_++J^i_-+J^i=0$ (implicitly imposed above) cannot be simultaneously strongly imposed as they do not form a Lie algebra. One has to impose them weakly and there are two possibilities.

In order to see this let us exploit the fact that there is a strict analogy with the way the simplicity constraints are imposed in the EPRL-FK model \cite{FK, EPRL}. Observe first that equation (\ref{EPRL}) has the very same form of the linear simplicity constraints of the EPRL-FK models where the role of the Immirzi parameter is here played by $\alpha$. 

The first possibility of weak imposition consists of
taking \be\label{op1} j_{\pm}=(1\pm\alpha) j/2\ee  implying \be
 j=j_++j_-.
 \ee
It can be checked that this choice is
consistent with alpha as given in (\ref{alfa}). With this then one can check that for an admissible state $|\psi\rangle$ one has
$$C^2 |\psi\rangle=\hbar^2(1-\alpha^2)j |\psi\rangle,$$
which vanishes in the (semiclassical) limit $\hbar\to 0$,  $j\to \infty$ with $\hbar j$ kept constant. Moreover, one has that
\be\label{weak} \langle\phi |C^i|\psi\rangle=0\ee
for arbitrary pairs of admissible states.
In other words, in this first possibility the constraint $C^i$ are satisfied strongly in the semiclassical limit, and weakly in the sense of matrix elements in general.

The second possibility is not to impose the condition $(\ref{op1})$ and leave $j_{\pm}$ completely free and only constrained by the triangular inequalities with $j$.
In that case it has been shown \cite{SF} that $(\ref{weak})$ is still satisfied. This second possibility is still compatible with the classical limit but is weaker than the previous 
one.
\vskip 0.5cm
\noindent{\bf{\em Entropy calculation A}}

In this case we impose condition (\ref{op1}) and hence $j=j_++j_-$.
Such a state characterizes a
black hole of macroscopic area
$$
a \; = \; \frac{1}{2} \, \sum_{\ell=1}^p\sqrt{(d_\ell^++d_\ell^--1)(d_\ell^++d_\ell^- +1)}
$$
in unit of $\ell_p^2$. 
As a consequence, the number of microstates $N(a)$ associated to a distorted black hole of area $a$
is given by the formula:
\ba
N(a) =\sum_{p=0}^\infty \sum_{d^{\pm}_1,\cdots,d^{\pm}_p}  \delta(a-\frac{\sum_{\ell=1}^p\sqrt{(d_\ell^++d_\ell^--1)(d_\ell^++d_\ell^-+1)}}{2}) \, \widetilde{N}_{k}({\bf j^+})\widetilde{N}_{k}({\bf j^-}),
\ea
where $d^\pm_\ell=2j_\ell^\pm+1$.
Following the steps of the previous Section, we introduce the Laplace transform $\tilde{N}(s)$ of the number of states $N(a)$.
It is given by:
\begin{eqnarray*}
\widetilde{N}(s)&=& \sum_{p=0}^\infty  \sum^{k+1}_{{\bf d^+},{\bf d^-}} \, \int_{0}^\pi d\theta^+ \, \mu_k(\theta^+)
\int_{0}^\pi d\theta^- \, \mu_k(\theta^-) \n\\
&\cdot&\left( \prod_{\ell=1}^p \frac{\sin(\frac{d^+_\ell \theta^+}{2})}{\sin \frac{\theta^+}{2}}
\frac{\sin(\frac{d^-_\ell \theta^-}{2})}{\sin \frac{\theta^-}{2}} e^{-\frac{s}{2} \sqrt{(d_\ell^++d_\ell^- -1)(d_\ell^++d_\ell^- +1)}}\right),
\end{eqnarray*}
where the sums run over the families 
${\bf d}^\pm\!=\!(d_1^\pm,\cdots,d_p^\pm)$
of representations dimensions.
Following the same strategy as in the spherically symmetric case, the previous expression can be simplified as follows:
\begin{eqnarray*}
\widetilde{N}(s) &=& \sum_{p=0}^\infty \int_0^\pi d\theta^+ \,\mu_k(\theta^+) \int_{0}^\pi d\theta^- \, \mu_k(\theta^-)\nonumber\\
&\cdot&\left( \sum^{k+1}_{d^+,d^-=1}  \frac{\sin (\frac{d^+\theta^+}{2})}{\sin \frac{\theta^+}{2}}
\frac{\sin (\frac{d^- \theta^-}{2})}{\sin \frac{\theta^-}{2}} \, e^{-\frac{s}{2} \sqrt{(d_\ell^++d_\ell^- -1)(d_\ell^++d_\ell^- +1)}}\right)^p\\
&=& \int_0^\pi d\theta^+ \int_{0}^\pi d\theta^- \; \frac{\mu_k(\theta^+)\mu_k(\theta^-)}{D^A_k(\theta^+,\theta^-;s)},
\end{eqnarray*}
with
\begin{eqnarray}
 D^A_k(\theta^+,\theta^-;s) \; = \; 1 -\sum_{d^\pm=0}^k  \, \frac{\sin (\frac{(d^++1)\theta^+}{2})}{\sin \frac{\theta^+}{2}}
\frac{\sin (\frac{(d^-+1) \theta^-}{2})}{\sin \frac{\theta^-}{2}} \, 
e^{-\frac{s}{2} \sqrt{(d_\ell^++d_\ell^- +1)(d_\ell^++d_\ell^- +3)}} \,.\label{DkA}
\end{eqnarray}
In the definition of $D^A_k$, the sums run over  $d^\pm \in [0,k]$, which are related to the spin variables $j^\pm$
by the relation $d^\pm=2j^\pm$ due to the changing of variables. 

\begin{figure}[ht]
\centering
\includegraphics[scale=1]{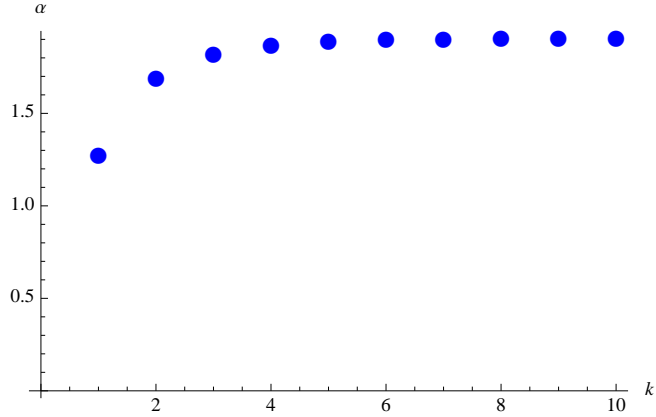}
\caption{In the figure, we have plotted the values of the exponent $\alpha $ as function of $k\in \mathbb N$ for the first integers. The plot shows as, similarly to the spherically symmetric black hole counting, in the case of a weak imposition of the constraint $C^i=0$ through the relation $j=j_++j_-$, an asymptotic value for $\alpha$ is quickly reached as $k\geq 4$.}
\label{Fig:s-kDistorted1A}
\end{figure}

As in the spherically symmetric case, we conclude that the critical value of $s$ is the highest value for which $D^A_k$, viewed as
a function of the angles $\theta^\pm$, admits a zero. It is reached when $D^A_k$ admits one zero at $\theta^\pm=0$.
Therefore, the critical exponent $\alpha$ is the unique solution of the equation
\be\label{k-sA}
1 -  \sum^{k}_{d^\pm=0} (d^++1)(d^-+1) e^{-\frac{\alpha}{2} \sqrt{(d_\ell^++d_\ell^- +1)(d_\ell^++d_\ell^- +3)}} \; = \;0 \, .
\ee
The exponent $\alpha$ depends on $k$ as in the spherically symmetric case and the numerical solution has been plotted
in Figure \ref{Fig:s-kDistorted1A}.

{\bf
\noindent{\em Entropy calculation B} }

Let us now concentrate on the case of the weaker imposition of the constraint $C^i=0$, where all pairs of admissible states are taken into account. In this case, the black hole of macroscopic area is 
$$
a \; = \; \frac{1}{2} \, \sum_{\ell=1}^p\sqrt{(d_\ell-1)(d_\ell +1)}
$$
in unit of $\ell_p^2$. 
As a consequence, the number of microstates $N(a)$ associated to a distorted black hole of area $a$
is given by the formula:
\ba
N(a) & = &\sum_{p=0}^\infty \sum_{d_1,\cdots,d_p}  \delta(a-\frac{\sum_{\ell=1}^p\sqrt{(d_\ell-1)(d_\ell +1)}}{2}) \nonumber \\
&\cdot& \sum^{k+1}_{d^\pm_1,\cdots,d^\pm_p=1} \left(\prod_{\ell=1}^pY(j_\ell,j_\ell^+,j_\ell^-)\right) \, \widetilde{N}_{k}({\bf j^+})\widetilde{N}_{k}({\bf j^-}),
\ea
where $d^\pm_\ell=2j_\ell^\pm+1$ and we recall that, in order to implement the admissibility condition, $Y(j_\ell,j_\ell^+,j^-_\ell)=1$ if $(j_\ell,j^+_\ell,j^-_\ell)$ satisfy
the triangular inequality, it vanishes otherwise.

Following the steps of the previous case A, we introduce the Laplace transform $\tilde{N}(s)$ of the number of states $N(a)$.
It is given by:
\begin{eqnarray*}
\widetilde{N}(s) & = & \sum_{p=0}^\infty \sum_{\bf d} \sum^{k+1}_{{\bf d^+},{\bf d^-}} \, \int_{0}^\pi d\theta^+ \, \mu_k(\theta^+)
\int_{0}^\pi d\theta^- \, \mu_k(\theta^-) \nonumber\\
&\cdot&\left( \prod_{\ell=1}^p Y(j_\ell,j_\ell^+,j^-_\ell) \, \frac{\sin(\frac{d^+_\ell \theta^+}{2})}{\sin \frac{\theta^+}{2}}
\frac{\sin(\frac{d^-_\ell \theta^-}{2})}{\sin \frac{\theta^-}{2}} e^{-\frac{s}{2} \sqrt{(d_\ell -1)(d_\ell +1)}}\right)\nonumber\,.
\end{eqnarray*}
Following the same strategy as in the case A, the previous expression can be simplified as follows:
\begin{eqnarray*}
\widetilde{N}(s) &=& \sum_{p=0}^\infty \int_0^\pi d\theta^+ \,\mu_k(\theta^+) \int_{0}^\pi d\theta^- \, \mu_k(\theta^-)\nonumber\\
&\cdot& \left( \sum_d \sum^{k+1}_{d^+,d^-=1} Y(j_\ell,j_\ell^+,j^-_\ell) \frac{\sin (\frac{d^+\theta^+}{2})}{\sin \frac{\theta^+}{2}}
\frac{\sin (\frac{d^- \theta^-}{2})}{\sin \frac{\theta^-}{2}} \, e^{-\frac{s}{2} \sqrt{(d-1)(d+1)}}\right)^p\\
&=& \int_0^\pi d\theta^+ \int_{0}^\pi d\theta^- \; \frac{\mu_k(\theta^+)\mu_k(\theta^-)}{D^B_k(\theta^+,\theta^-;s)},
\end{eqnarray*}
with now
\begin{eqnarray}
 D^B_k(\theta^+,\theta^-;s) \; = \; 1 -\sum_{d} \sum_{d^\pm=0}^k Y(j_\ell,j_\ell^+,j^-_\ell) \, \frac{\sin (\frac{(d^++1)\theta^+}{2})}{\sin \frac{\theta^+}{2}}
\frac{\sin (\frac{(d^-+1) \theta^-}{2})}{\sin \frac{\theta^-}{2}} \, e^{-\frac{s}{2} \sqrt{d(d+2)}} \,.\label{Dk}
\end{eqnarray}
Again, the spins variables $j^\pm$
are related to the sums variables by $d^\pm=2j^\pm$ due to the changing of variables. The variable $d$ is also related to the spin variable
$j$ by $d=2j$.

\begin{figure}[ht]
\centering
\includegraphics[scale=1]{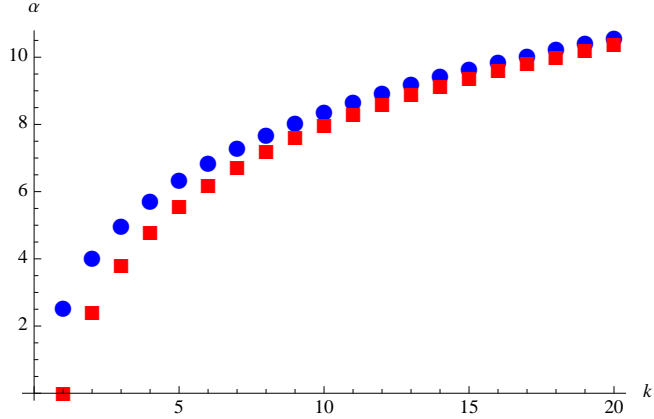}
\caption{In the plot, the circles indicate the values of the exponent $\alpha$ as function of $k\in \mathbb N$ for the first integers; the squares represent values of the function $c\log{k}$ for the same values of $k$, where $c=2\sqrt{3}+O(1/k)$.}
\label{Fig:s-kDistorted1B}
\end{figure}

Similarly to the previous cases, we conclude that the critical value of $s$ is the highest value for which $D^B_k$, viewed as
a function of the angles $\theta^\pm$, admits a zero. It is reached when $D^B_k$ admits one zero at $\theta^\pm=0$.
Therefore, the critical exponent $\alpha$ is the unique solution of the equation
\be\label{k-s}
1 - \sum_d \sum^{k}_{d^\pm=0} Y(j_\ell,j_\ell^+,j^-_\ell) \, (d^++1)(d^-+1) e^{-\frac{\alpha}{2} \sqrt{d(d+2)}} \; = \;0 \, .
\ee
The exponent $\alpha$ again depends on $k$ and the numerical solutions has been plotted
in Figure \ref{Fig:s-kDistorted1B}.

To have a more physical intuition of the behavior of $\alpha$ as a function of $k$, let us assume that all the spins $j$
are fixed to $1/2$: this means that the edges of the spin-network in the bulk which intersect the black hole surface
are colored by $1/2$-spins. This assumption will give us the ``shape'' of the function $\alpha(k)$ for large values of $k$ since
the main contributions to the entropy come from small values of the bulk spin $j_\ell$. Let us call $\alpha_{1/2}$ the value
of $\alpha$ where only $j=1/2$ spins contribute and $\alpha_{1/2}$ satisfies:
\be
\sum_{d^+=0}^k \sum_{d^-=d^+-1}^{d^++1} (d^++1)(d^-+1) e^{-\alpha_{1/2}\frac{\sqrt{3}}{2}} \;= \; 1 \,.
\ee
A straightforward calculation leads to the following expression relating $\alpha_{1/2}$ and $k$:
\be
e^{\alpha_{1/2}\frac{\sqrt{3}}{2}} \; = \; 3 \sum_{n=1}^{k+1} n^2 \; = \;  \frac{(k+1)(k+2)(2k+3)}{2} \,.
\ee
As a consequence, in the limit where $k$ is large
$$\alpha_{1/2} \sim 2\sqrt{3} \, \log k$$
which means that $\alpha$ grows logarithmically with $k$.
Further evidence of this behavior of $\alpha$ is given by the numerical solution of eq. (\ref{k-s}) plotted above (Figure \ref{Fig:s-kDistorted1B})
where all spins are taken into account and not only $1/2$ spins. Note that the numerical value $c$ in the asymptotic formula
$\alpha\sim c\log k$ is such that $c=2\sqrt{3}+O(1/k)$ as expected.

\medskip

Let us now concentrate on the sub-leading corrections. We proceed exactly as in the spherically symmetric case:
we first evaluate $\tilde{N}(s)$ at the vicinity of the critical point $\alpha$, i.e. $s=\alpha + \varepsilon$ for
a small $\varepsilon$; we can then study the singularities of the generalized Laplace transform $\widetilde{N}(\alpha,t)$ evaluated at the critical value through the relation (\ref{generalizedLaplace}), by expanding the integrand $f_\varepsilon(\theta^+,\theta^-)$ around $\theta^+=0=\theta^-$; finally, we look at the maximal value
of $t_c$ for which $\tilde{N}(\alpha,t)$ is well-defined. The critical exponent $\beta$ is then given by $\beta=-t_c-1$. More precisely, in the distorted case we have
\begin{eqnarray*}
&&\frac{\partial^{t} f_\varepsilon(\theta^+,\theta^-) \vert_{\varepsilon=0}}{\partial \varepsilon^t} 
\sim   \frac{\theta^{+2}\theta^{-2}}{(\theta^{+2(t+1)}+\theta^{-2(t+1)})} =\frac{\rho^4}{\rho^{2(t+1)}}\frac{\sin^2({\varphi})\cos^2(\varphi)}{(\sin^{2(t+1)}(\varphi)+\cos^{2(t+1)}(\varphi))},
\end{eqnarray*}
where in the last equality we have changed to polar coordinates $\theta^+=\rho \sin(\varphi)$, $\theta^-=\rho \cos(\varphi)$.
From the previous equation we get that $$\tilde{N}(\alpha, -t)\simeq \int d\rho  \,\rho^{-2t+3}.$$ Consequently, $\tilde{N}(\alpha, t)$ is singular when $t > 2$, in this case the critical value of $t$ is $t_c=2$.
Therefore, the critical exponent now is $\beta=-t_c-1=-3$ and the asymptotic expansion reads:
\be
N(a) \; \sim \; e^{\alpha a} \, a^{-3}   \;,
\ee
where $\alpha$ is given in Figure \ref{Fig:s-kDistorted1A} or \ref{Fig:s-kDistorted1B} according to the prescription that defines the allowed states.
From the previous expression for the Laplace transform, we see that, in the distorted case, the constant factor in front of the logarithmic corrections becomes $3$.

\subsection*{Conclusion}
This paper has been devoted to the calculation of leading and sub-leading terms of the $SU(2)$
black hole entropy in Loop Quantum Gravity when the black hole is spherically symmetric \cite{su21,su22}
and when it is distorted \cite{su23}.  To reach this aim, we derived first, by means of the recoupling theory of the quantum group $U_q(su(2))$, a new integral formula, resulting to be very useful, for the dimension of the
Hilbert space of $SU(2)$ Chern-Simons theory on a punctured two-sphere, which enters the definition of the Hilbert space of the $SU(2)$ spherically symmetric and distorted black hole as derived in \cite{su22,su23}. 
Successively, we revised the technique of the Laplace transform method, exposed in detail in \cite{counting3} and firstly introduced in \cite{counting1}, to study the asymptotic behavior (in the large area limit) of the entropy associated to these two statistic mechanical ensembles.

The entropy of a $SU(2)$ spherically symmetric black hole has been already studied in \cite{counting3} when the
level $k$ is infinite. Here, following a paradigm-shift introduced in \cite{su23}, we considered the case where the level
$k$ of the Chern-Simons theory and the macroscopic area of the black hole $a$ are independent variables and we studied the effect of a 
finite $k$. We showed that, if one takes into account the
finiteness of the level, the entropy of type I isolated horizons is not modified at least up to the subleading corrections, therefore recovering the results of \cite{counting3}. Moreover, the critical exponent of the leading order $\alpha$, which is now a function of the level $k$, reaches fast an asymptotic value for large $k$, as shown in the plot in Figure 6.

Concerning the entropy of a distorted $SU(2)$ black hole, this is something which has never been studied before. In this case, for each puncture coming from the bulk, there are two punctures associated to it on the horizon and the Hilbert  space becomes now the direct product of two $SU(2)$ Chern-Simons Hilbert  spaces with same level $k$ \cite{su23}. The $SU(2)$ symmetry is implemented by the insertion of an intertwiner between the three punctures (one from the bulk and two from the horizon), therefore, the area constraint still plays an important role. Using the techniques developed for the spherically symmetric case, we have performed the counting of the enlarged Hilbert space number of states and shown that the entropy is again proportional to the horizon area to the leading order. In the distorted case, one can distinguish two different models according to the way  second class constraints are imposed weakly. In the strongest imposition of the constraints the results do not differ in a qualitative sense from those obtained in the spherically symmetric case. However, if the second class constraints are only imposed weakly, in the Gupta-Bleurer sense, then the critical exponent $\alpha$ does not go to an asymptotic value for increasing values of the level $k$ but grows logarithmically with it, as shown in Figure \ref{Fig:s-kDistorted1B}. In that sense, our model is
consistent with ``any'' value of the Immirzi parameter.

\subsubsection*{Acknowledgments}
This project was partially supported by the ANR. D.P. was supported by the {\em Marie Curie} EU-NCG network.


\begin{thebibliography}{100}

\bibitem{lqg}
A.~Ashtekar and J.~Lewandowski,
  Class.\ Quant.\ Grav.\  {\bf 21}, R53 (2004)
  [arXiv:gr-qc/0404018].
  
\newblock  C.~Rovelli,
  ``Quantum Gravity,''
{\it  Cambridge, UK: Univ. Pr. (2004) 455 p}.

\newblock T.~Thiemann,
  ``Modern canonical quantum general relativity,''
  {\it  Cambridge, UK: Univ. Pr. (2007) 819 p}.
  
\newblock A.~Perez,
  ``Introduction to loop quantum gravity and spin foams,''
  [arXiv:gr-qc/0409061].

\bibitem{ih}
  A.~Ashtekar and B.~Krishnan,
  ``Isolated and dynamical horizons and their applications,''
  Living Rev.\ Rel.\  {\bf 7} (2004) 10
  [arXiv:gr-qc/0407042].


\bibitem{ab}
  A.~Ashtekar, J.~C.~Baez and K.~Krasnov,
  Adv.\ Theor.\ Math.\ Phys.\  {\bf 4} (2000) 1
  [arXiv:gr-qc/0005126].

\bibitem{jon}
  C.~Beetle and J.~Engle,
  ``Generic isolated horizons in loop quantum gravity,''
  Class.\ Quant.\ Grav.\  {\bf 27} (2010) 235024
  [arXiv:1007.2768 [gr-qc]].


\bibitem{su21}
  J.~Engle, A.~Perez and K.~Noui,
  ``Black hole entropy and SU(2) Chern-Simons theory,''
  Phys.\ Rev.\ Lett.\  {\bf 105} (2010) 031302
  [arXiv:0905.3168 [gr-qc]].

\bibitem{su22}
  J.~Engle, K.~Noui, A.~Perez and D.~Pranzetti,
  ``Black hole entropy from an SU(2)-invariant formulation of Type I isolated
  horizons,''
  Phys.\ Rev.\  D {\bf 82} (2010) 044050
  [arXiv:1006.0634 [gr-qc]].

\bibitem{su23}
  A.~Perez and D.~Pranzetti,
  ``Static isolated horizons: SU(2) invariant phase space, quantization, and
  black hole entropy,''
  arXiv:1011.2961 [gr-qc].

\bibitem{counting1}   K.~A.~Meissner,
  ``Black hole entropy in loop quantum gravity,''
  Class.\ Quant.\ Grav.\  {\bf 21} (2004) 5245
  [arXiv:gr-qc/0407052].
  
\newblock M.~Domagala and J.~Lewandowski,
  ``Black hole entropy from quantum geometry,''
  Class.\ Quant.\ Grav.\  {\bf 21} (2004) 5233
  [arXiv:gr-qc/0407051].
  

\bibitem{counting2}
A.~Ghosh and P.~Mitra,
  ``Fine-grained state counting for black holes in loop quantum gravity,''
  Phys.\ Rev.\ Lett.\  {\bf 102} (2009) 141302
  [arXiv:0809.4170 [gr-qc]].
  
\newblock  A.~Ghosh and P.~Mitra,
  ``Counting black hole microscopic states in loop quantum gravity,''
  Phys.\ Rev.\  D {\bf 74} (2006) 064026
  [arXiv:hep-th/0605125].
  
\newblock   A.~Ghosh and P.~Mitra,
  ``Counting of black hole microstates,''
  Indian J.\ Phys.\  {\bf 80} (2006) 867
  [arXiv:gr-qc/0603029].

\newblock A.~Ghosh and P.~Mitra,
  ``An improved lower bound on black hole entropy in the quantum geometry
  approach,''
  Phys.\ Lett.\  B {\bf 616} (2005) 114
  [arXiv:gr-qc/0411035].
 
\newblock  A.~Ghosh and P.~Mitra,
  ``A bound on the log correction to the black hole area law,''
  Phys.\ Rev.\  D {\bf 71} (2005) 027502
  [arXiv:gr-qc/0401070].

\bibitem{counting3}
I.~Agullo, J.~Fernando Barbero, E.~F.~Borja, J.~Diaz-Polo and E.~J.~S.~Villasenor,
  ``Detailed black hole state counting in loop quantum gravity,''
  Phys.\ Rev.\  D {\bf 82} (2010) 084029
  [arXiv:1101.3660 [gr-qc]].
  
\newblock I.~Agullo, G.~J.~Fernando Barbero, E.~F.~Borja, J.~Diaz-Polo and E.~J.~S.~Villasenor,
  ``The combinatorics of the SU(2) black hole entropy in loop quantum
  gravity,''
  Phys.\ Rev.\  D {\bf 80} (2009) 084006
  [arXiv:0906.4529 [gr-qc]].
  
\newblock   J.~F.~Barbero G. and E.~J.~S.~Villasenor,
  ``On the computation of black hole entropy in loop quantum gravity,''
  Class.\ Quant.\ Grav.\  {\bf 26} (2009) 035017
  [arXiv:0810.1599 [gr-qc]].
  
\newblock  J.~F.~Barbero G. and E.~J.~S.~Villasenor,
  ``Generating functions for black hole entropy in Loop Quantum Gravity,''
  Phys.\ Rev.\  D {\bf 77} (2008) 121502
  [arXiv:0804.4784 [gr-qc]].
  
\newblock I.~Agullo, J.~F.~Barbero G., J.~Diaz-Polo, E.~Fernandez-Borja and E.~J.~S.~Villasenor,
  ``Black hole state counting in LQG: A number theoretical approach,''
  Phys.\ Rev.\ Lett.\  {\bf 100} (2008) 211301
  [arXiv:0802.4077 [gr-qc]].
  
\bibitem{hanno}  H.~Sahlmann,
  ``Entropy calculation for a toy black hole,''
  Class.\ Quant.\ Grav.\  {\bf 25} (2008) 055004
  [arXiv:0709.0076 [gr-qc]].
  
\newblock H.~Sahlmann,
  ``Toward explaining black hole entropy quantization in loop quantum
  gravity,''
  Phys.\ Rev.\  D {\bf 76} (2007) 104050
  [arXiv:0709.2433 [gr-qc]].


\bibitem{kaulma}
  R.~K.~Kaul and P.~Majumdar,
  ``Quantum black hole entropy,''
  Phys.\ Lett.\  B {\bf 439} (1998) 267
  [arXiv:gr-qc/9801080].
  
\newblock  R.~K.~Kaul and P.~Majumdar,
  ``Logarithmic correction to the Bekenstein-Hawking entropy,''
  Phys.\ Rev.\ Lett.\  {\bf 84} (2000) 5255
  [arXiv:gr-qc/0002040].
  
 
\newblock  S.~Das, R.~K.~Kaul and P.~Majumdar,
  ``A new holographic entropy bound from quantum geometry,''
  Phys.\ Rev.\  D {\bf 63} (2001) 044019
  [arXiv:hep-th/0006211].

\bibitem{livi}
  E.~R.~Livine and D.~R.~Terno,
  ``Quantum black holes: Entropy and entanglement on the horizon,''
  Nucl.\ Phys.\  B {\bf 741} (2006) 131
  [arXiv:gr-qc/0508085].
  
\newblock   L.~Freidel and E.~R.~Livine,
  ``The Fine Structure of SU(2) Intertwiners from U(N) Representations,''
  J.\ Math.\ Phys.\  {\bf 51} (2010) 082502
  [arXiv:0911.3553 [gr-qc]].
  
\newblock    E.~R.~Livine and D.~R.~Terno,
  ``The entropic boundary law in BF theory,''
  Nucl.\ Phys.\  B {\bf 806} (2009) 715
  [arXiv:0805.2536 [gr-qc]].
  
\newblock  E.~R.~Livine and D.~R.~Terno,
  ``Bulk Entropy in Loop Quantum Gravity,''
  Nucl.\ Phys.\  B {\bf 794} (2008) 138
  [arXiv:0706.0985 [gr-qc]].
  



\bibitem{AT}
Achucarro A., Townsend P.,
{``A Chern--Simons action for three-dimensional anti-de Sitter supergravity theories''},
{\it Phys.Lett. B 180}, 85--100 (1986).

\bibitem{Witten1}
Witten E.,
``2+1 dimensional gravity as an exactly soluble system",
{\it Nucl. Phys. B 311},  46--78 (1988).

\bibitem{Witten2}
Witten E.,
``Quantum field theory and the Jones polynomial",
{\it Commun. Math. Phys.  121}, 351 (1989).


\bibitem{CK}
C.~Meusburger and K.~Noui,
  ``Combinatorial quantisation of the Euclidean torus universe,''
  Nucl.\ Phys.\  B {\bf 841}, 463 (2010)
  [arXiv:1007.4615 [gr-qc]].

\bibitem{CP}
Chary V., Pressley A.,
``A guide to quantum groups'',
{\it Cambridge University Press} (1994).

\bibitem{Verlinde}
Verlinde E.,
``Fusion Rules and Modular Transformations in 2D Conformal Field Theory'',
{\it Nucl.Phys. B 300} 360 (1988).


\bibitem{EPRL}
  J.~Engle, E.~Livine, R.~Pereira and C.~Rovelli,
  ``LQG vertex with finite Immirzi parameter,''
  Nucl.\ Phys.\  B {\bf 799} (2008) 136
  [arXiv:0711.0146 [gr-qc]].



\bibitem{FK}
  L.~Freidel and K.~Krasnov,
  ``A New Spin Foam Model for 4d Gravity,''
  Class.\ Quant.\ Grav.\  {\bf 25} (2008) 125018
  [arXiv:0708.1595 [gr-qc]].


\bibitem{SF}
  Y.~Ding, M.~Han and C.~Rovelli,
  ``Generalized Spinfoams'';
 [arXiv:gr-qc/10112149].



	
\end{thebibliography}
\end{document}